\documentclass[paper,12pt,nofootinbib,notitlepage]{revtex4-1}
\usepackage{graphicx}
\usepackage{amsmath}
\usepackage{amssymb}
\usepackage{dcolumn}% Align table columns on decimal point
\usepackage{bm}% bold math
\usepackage{color}
\usepackage{dsfont}
\usepackage{feynmp}

\newcommand \beq{\begin{eqnarray}}
\newcommand \eeq{\end{eqnarray}}

\begin{document}
\unitlength=1mm
\allowdisplaybreaks

\title{The scalar sunset diagram at finite temperature\\ with imaginary square masses}

\author{Duifje Maria van Egmond}
\affiliation{Universidade do Estado do Rio de Janeiro, Instituto de F\'isica --- Departamento de F\'isica
Te\'orica\\ Rua Sao Francisco Xavier 524, 20550-013, Maracana, Rio de Janeiro, Brasil}

\author{Urko Reinosa}
\affiliation{Centre de Physique Th\'eorique, CNRS, Ecole polytechnique, IP Paris, F-91128 Palaiseau, France.}

\date{\today}

\begin{abstract}
We evaluate the finite temperature scalar sunset diagram with imaginary square masses, that appears in the Gribov-Zwanziger approach to Yang-Mills (YM) theory beyond one-loop order. Since YM theory at finite temperature is governed by center-symmetry and the Polyakov loop, we also include the possibility of a constant temporal background gauge field in the form of color-dependent imaginary chemical potentials.
\end{abstract}

\maketitle

%%%%%
\section{Introduction}
In recent years, much valuable progress has been made towards the understanding of non-abelian gauge theories at finite temperature using background field gauge (BFG) methods \cite{Abbott:1980hw,Abbott:1981ke} in the Landau-DeWitt gauge, in combination with several functional methods \cite{Braun:2007bx,Braun:2009gm,Braun:2010cy,Fischer:2009wc,Fischer:2009gk,Fischer:2011mz,Fischer:2013eca,Fischer:2014vxa,Reinhardt:2012qe,Reinhardt:2013iia,Quandt:2016ykm,Reinhardt:2017pyr}. On the one hand, BFG methods provide an efficient way to describe the confinement/deconfinement order parameter (the Polyakov loop or any of its proxies \cite{Braun:2007bx}) because the related center symmetry is explicit at the quantum level and is easily maintained in approximation schemes \cite{Herbst:2015ona,Reinosa:2015gxn,Reinosa_HDR}. On the other hand, functional methods provide a method of choice when investigating infrared, non-perturbative properties of non-abelian theories \cite{Fu:2019hdw}.

However, most functional approaches take as a starting point the usual Faddeev-Popov version of the gauge fixing which is known to be a valid description of non-abelian gauge theories at high energies but which is also expected to be modified in the infrared due to the influence of Gribov copies \cite{Gribov77}. It is then an interesting question whether a complete gauge-fixing procedure in the infrared (IR) regime could capture some genuine non-perturbative effects, beyond those that are captured by the infinite hierarchies of equations considered in functional methods. Even more, it has been suggested that a resolution of the IR gauge-fixing may open the way to a new perturbative perspective on certain aspects of the infrared dynamics of non-abelian gauge fields \cite{Tissier:2010ts,Tissier:2011ey}.

Several models have been put forward in order to implement the BFG formalism in the Landau-deWitt gauge while restricting the number of Gribov copies. In Refs.~\cite{Reinosa:2014ooa, Reinosa:2014zta,Reinosa:2015gxn,Reinosa:2015oua,Maelger:2017amh}, the formalism was used within the Curci Ferrari (CF) model \cite{Curci76} to compute the background potential and Polyakov loop up to two-loop order, both in pure Yang-Mills theories and in heavy-quark QCD. It was argued that this model could be part of a complete gauge-fixing in the Landau gauge, since a CF gluon mass term may arise after the Gribov copies have been accounted for via an averaging procedure \cite{Serreau:2012cg}, see also Ref.~\cite{Tissier:2017fqf} for a related discussion in a different gauge. One salient feature of the results obtained within the CF model is that, not only various aspects of the phase structure are already accounted for at leading one-loop order, but the two-loop corrections turn out to be small and tend to improve the results, supporting the existence of a ``perturbative way'' lurking behind the gauge-fixing problem.

A more explicit way to account for Gribov copies is the Gribov-Zwanziger (GZ) method \cite{Zwanziger:1989mf,Vandersickel:2012tz,Dudal08}. At the cost of introducing some new fields, the functional integral is restricted to a region that contains less Gribov copies, the so-called Gribov region. In Ref.~\cite{Canfora:2015yia}, a GZ type action for the Landau-DeWitt gauge was proposed, but it was later established in Ref.~\cite{Dudal:2017jfw} that this model is not invariant under background gauge transformations and an alternative proposal was made where both background gauge invariance and BRST symmetry are manifest. It is however not clear how to extend this proposal at finite temperature while maintaining the background gauge invariance. Here, we will follow the framework of Ref.~\cite{Kroff:2018ncl}, where BRST symmetry is sacrificed (just as in the CF model) to establish a background gauge invariant GZ type model that is easy to implement at finite temperature.

In Ref.~\cite{Kroff:2018ncl}, the one-loop background potential and the Polyakov loop up to first order were determined within this model and in Ref.~\cite{Maelger:2018vow} these calculations were extended to the case of QCD with heavy quarks, leading to the the best agreement to date with the available lattice data regarding the description of the upper boundary line in the so-called Columbia plot. A natural question is whether these promising results at leading order resist the inclusion of higher order corrections, which would support similar results within the CF approach.

The present work is a modest contribution towards this goal: we address the calculation of the scalar sunset diagram and the mass derivatives thereof that appear in the  two-loop background potential in the model at finite temperature. This potential puts forward new challenges because imaginary square masses appear with the introduction of the auxiliary fields needed to localize the GZ action. Indeed, the tree-level gluon propagator in the GZ model reads 
\beq
G_0(Q)=\frac{Q^2}{Q^4+\gamma^4}={\rm Re}\,\frac{1}{Q^2+i\gamma^2}\,,\nonumber
\eeq
with $\gamma$ the Gribov parameter. Though the existence of imaginary masses in the GZ model is a well-known fact, to our knowledge there is no literature on the proper handling of imaginary masses in higher-order loop calculations at finite temperature. The full calculation of the two-loop potential in the GZ model as well as the Polyakov loop will be treated in a different work \cite{vEgmond}. For related work at zero temperature, see \cite{Gracey:2005cx,Ford:2009ar,Gracey:2010cg}.
 
A convenient tool to make sense of the finite temperature contributions to the potential is  {\it thermal splitting} which is commonly used in calculations that involve Matsubara sums \cite{Andersen:2000zn,Blaizot:2004bg}. By decomposing sum-integrals according to the number of thermal factors, UV divergences become much easier to handle. Moreover, we can separate a vacuum piece, which will equal the zero-temperature contribution. The vacuum two-loop sunset amplitude for real masses was calculated in \cite{Caffo:1998du} and the finite temperature contributions have been known for a long time \cite{Parwani:1991gq}, with a recent generalization in the presence of the Polyakov loop \cite{Marko:2010cd,Reinosa:2014zta,Reinosa:2015gxn}. Part of this work will therefore be an extension of these results to the case of imaginary square masses. We do not aim at a full generalization, however, instead limiting ourselves to the cases that appear in the two-loop calculation in the GZ model \cite{vEgmond}.

This work is organized as follows. In Sec. \ref{sec:tadpole}, we look at the scalar tadpole sum-integral as a pedagogical introduction to the techniques that will be used to deal with the sunset sum-integral. In particular, we introduce the spectral representation and give a first trivial example of thermal splitting. In Sec. \ref{sec:sunset}, we look at the scalar sunset sum-integral. In Sec. \ref{sec:mass_derivatives}, we investigate the relevant mass derivatives of the sunset sum-integrals and their respective thermal splittings that are also needed for the evaluation of the GZ potential at two-loop order. More technical details are gathered in the Appendices.

%%%%%
\section{The scalar tadpole as a simple example}\label{sec:tadpole}
In what follows, we denote Euclidean momenta by capital letters $Q$, $K$, $L$, $\dots$ Each of these comprises a bosonic Matsubara frequency $\smash{\omega_n\equiv 2\pi T n}$, with $\smash{n\in\mathds{Z}}$, and a spatial momentum ${\bf q}$\,, with $\smash{q\equiv|{\bf q}|}$. Integration over Euclidean momenta is encoded in sum-integrals, which we keep denoting, however, as standard integrals for simplicity:
\beq
\int_Q f(Q)\equiv T\sum_{n\in\mathds{Z}} \int_{\bf q} f(\omega_n;\bf{q})\,.
\eeq
We work in dimensional regularization, meaning that the integral over spatial momenta corresponds to
\beq
\int_{\bf{q}} \equiv \mu^{2\epsilon}\int\frac{d^{d-1}q}{(2\pi)^{d-1}}\,,
\eeq
with $\smash{d=4-2\epsilon}$.

In the context of Yang-Mills theory at finite temperature, it is crucial to take into account the order parameter for the confinement/deconfinement transition also known as the Polyakov loop $\ell$, or, equivalently, the corresponding constant, temporal and diagonal gluonic background $\bar A_0$ such that $\ell\propto{\rm tr}\exp\{i g\beta\bar A_0\}$, with $\smash{\beta\equiv1/T}$ the inverse temperature. In this situation, the Matsubara frequencies are shifted by a color-dependent imaginary chemical potential, $\smash{\omega_n\to\omega_n^\kappa\equiv \omega_n+\hat{r}^j\kappa^j}$, where the $\hat{r}^j$ denote the components of $\bar A_0$ along the diagonal part $\{t^j\}$ of the $su(N)$ algebra, $\smash{\bar A_0=\hat r^j t^j}$, while the $\kappa^j$ denote the weights of the adjoint representation, that arise as one diagonalizes the adjoint action of all the $t^j$: $[t^j,t^\kappa]=\kappa^j t^\kappa$. For the present paper, we do not need to know more about the precise way the Polyakov loop appears in explicit calculations. In what follows, we denote by $\smash{Q_\kappa=(\omega_n^\kappa;{\bf q})}$ the shifted Euclidean momentum and we also introduce the notation $\hat r\cdot\kappa\equiv\hat{r}^j\kappa^j$.\\

In this first section, as a pedagogical example, we treat the scalar tadpole sum-integral
\beq\label{eq:tad}
J^\kappa_\alpha\equiv \int_Q G_\alpha(Q_\kappa)\,,
\eeq
with
\beq
G_\alpha(Q_\kappa)\equiv \frac{1}{Q_\kappa^2+\alpha}\,,
\eeq
assuming that the square mass $\alpha$ is purely imaginary. The procedure that follows might seem unnecessarily complicated for the evaluation of such a simple sum-integral. However, it introduces the basic ingredients that make the corresponding evaluation of the scalar sunset sum-integral in the next section much simpler.

%%%
\subsection{Spectral representation}
The first step is to evaluate the Matsubara sum in Eq.~(\ref{eq:tad}). To this purpose, we decompose the propagator as
\beq\label{eq:decomp}
G_\alpha(Q_\kappa)=\frac{1}{2\varepsilon_{q,\alpha}}\left[\frac{1}{\varepsilon_{q,\alpha}-i\omega_n^\kappa}-\frac{1}{-\varepsilon_{q,\alpha}-i\omega_n^\kappa}\right],
\eeq
with $\smash{\varepsilon_{q,\alpha}\equiv\sqrt{q^2+\alpha}}$. It proves useful to rewrite the previous identity in the form of a ``spectral representation''
\beq\label{eq:spectral}
G_\alpha(Q_\kappa)\equiv\tilde G_\alpha(i\omega_n^\kappa;{\bf q})=\int_{q_0}\frac{\rho_\alpha(q_0;{\bf q})}{q_0-i\omega_n^\kappa}\,,
\eeq
where $\smash{\int_{q_0}\equiv\int dq_0/(2\pi)}$ and
\beq
 \rho_\alpha(q_0;{\bf q})\equiv \frac{2\pi}{2\varepsilon_{q,\alpha}}\,\Big[\delta\!\left(q_0-\varepsilon_{q,\alpha}\right)-\delta\!\left(q_0+\varepsilon_{q,\alpha}\right)\Big]\equiv2\pi\,{\rm sign}(q_0)\delta\!\left(q^2_0-\varepsilon_{q,\alpha}^2\right).
\eeq
We mention that, in the presence of imaginary square masses, the notations $\int_{q_0}$ and $\delta\!\left(q_0\mp\varepsilon_{q,\alpha}\right)$ are understood as mere bookkeeping devices allowing to select the two complex energies $\pm\varepsilon_{q,\alpha}$. Similarly, ${\rm sign}(q_0)$ selects the corresponding sign in front of $\pm\varepsilon_{q,\alpha}$, and should therefore be understood as the sign of the real part of $q_0$.\footnote{The spectral representation can be given a rigorous meaning by defining the Dirac and sign distributions along the appropriate contour. We will not need these technicalities here though.} 

Using the spectral representation (\ref{eq:spectral}) in Eq.~(\ref{eq:tad}), we find
\beq
J^\kappa_\alpha & = & \int_{q_0;{\bf q}}\rho_\alpha(q_0;{\bf q})\,{\cal T}^\kappa(q_0)\,,\label{eq:8}
\eeq
with
\beq
{\cal T}^\kappa(q_0) & \equiv & T\sum_{n\in\mathds{Z}} \frac{1}{q_0-i\omega_n^\kappa}\,,
\eeq
a simple Matsubara sum. We stress that, even though $q_0$ takes complex values, it does not interfere with the Matsubara frequencies because its real part never vanishes. Using standard techniques for the evaluation of Matsubara sums, we then arrive at\footnote{The simple Matsubara sum considered here is not absolutely convergent. This means that, when applying the standard technique based on contour integration, one needs {\it a priori} to take into account a contribution from the contour at infinity. Fortunately, this contribution cancels upon integrating over $q_0$ in Eq.~(\ref{eq:8}), in line with the fact that the original Matsubara sum is absolutely convergent.}
\beq
{\cal T}^\kappa(q_0)=n_{q_0-i\hat r\cdot\kappa}\,,\label{eq:10}
\eeq 
with $n_x\equiv 1/(e^{x/T}-1)$ the Bose-Einstein distribution function. 

%%%
\subsection{Thermal splitting}
One problem with the expression above is that it involves thermal factors with energies whose real parts can be as negative as possible. In particular, this does not facilitate the extraction of UV divergences. To remedy this situation, we write
\beq
 n_{q_0-i\hat r\cdot\kappa} & = & -\theta(-q_0)+{\rm sign}(q_0)n_{|q_0|-i\,{\rm sign}(q_0)\hat r\cdot\kappa}\,,\label{eq:11}
\eeq
where ${\rm sign}(q_0)$ is to be understood as the sign of the real part of $q_0$, see above, $\theta(q_0)$ is equal to $1$ if the real part of $q_0$ is positive and zero otherwise, and $|q_0|=q_0\,{\rm sign}(q_0)$. Plugging Eq.~(\ref{eq:11}) into Eq.~(\ref{eq:10}) and then back into Eq.~(\ref{eq:8}), we arrive at the ``thermal splitting'' of the tadpole sum-integral: $J_\alpha^\kappa=J_\alpha(0n)+J_\alpha^\kappa(1n)$. Here, $J_\alpha(0n)$ denotes the pure vacuum contribution (no thermal factor), depending neither on the temperature nor on the background, while
\beq
J_\alpha^\kappa(1n)=\int_{q_0;{\bf q}}\sigma_\alpha^\kappa(q_0;{\bf q})\,,
\eeq
where $\sigma_\alpha^\kappa(q_0;{\bf q})\equiv\rho_\alpha(q_0;{\bf q})\,{\rm sign}(q_0)\,n_{|q_0|-i\,{\rm sign}(q_0)\hat r\cdot\kappa}$.

In the contribution with one thermal factor, one can perform the frequency integral. Moreover, because this contribution is UV finite, one can take the limit $d\to 4$ and evaluate the angular integral analytically. One obtains
\beq
J_\alpha^\kappa(1n)=\sum_{\sigma_\alpha}\int_{{\bf q}}\frac{n_{\varepsilon_{q,\alpha}-i\sigma _\alpha \hat r\cdot\kappa}}{2\varepsilon_{q,\alpha}}=\frac{1}{2\pi^2}\sum_{\sigma_\alpha}\int_0^\infty dq\,q^2\frac{n_{\varepsilon_{q,\alpha}-i\sigma _\alpha \hat r\cdot\kappa}}{2\varepsilon_{q,\alpha}}\,,
\eeq
where $\sigma_\alpha\in\{-1,+1\}$.  On the other hand, the vacuum contribution is conveniently computed by rewriting it as a standard $d$-dimensional Euclidean integral
\beq
J_\alpha(0n)=\int_Q^{T=0} G_\alpha(Q)\,.
\eeq
Seen as a function of a complex $\alpha$, this integral is analytic, with a branch cut for $\alpha \in \rm Re^-$. Therefore, its value for $\alpha$ imaginary can be obtained by analytic continuation of the known expression for $\alpha \in \rm Re^+$. We simply find
\beq
J_\alpha(0n)=-\frac{\alpha}{16\pi^2}\left[\frac{1}{\epsilon}+\ln\frac{\bar\mu^2}{\alpha}+1+{\cal O}(\epsilon)\right],
\eeq
where $\bar\mu^2\equiv4\pi\mu^2 e^{-\gamma}$ and $\gamma$ is the Euler-constant.

%%%%%
\section{Thermal splitting of the scalar sunset}\label{sec:sunset}
		\begin{figure}[h]
	\centering
	\includegraphics[width=8cm]{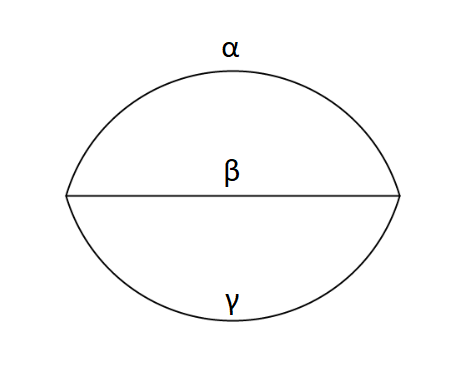}
	\caption{The two-loop sunrise graph for imaginary square masses $\alpha,\beta, \gamma$. The momenta and color charges are conserved at the vertices.}
	\label{Yw3}
\end{figure}
Using similar techniques, we now would like to evaluate the ($0$-leg) sunset sum-integral
\beq
S^{\kappa\lambda\tau}_{\alpha\beta\gamma}  \equiv  \int_{Q,K} G_\alpha(Q_\kappa)G_\beta(K_\lambda)G_\gamma(L_\tau)\,,\label{eq:sunset}
\eeq
where momentum and color conservation imply respectively $Q+K+L=0$ and $\kappa+\lambda+\tau=0$, see Fig. \ref{Yw3}.\footnote{These two identities can be conveniently combined into $Q_\kappa+K_\lambda+L_\tau=0$ \cite{Reinosa:2015gxn}.} We consider the case where the square masses $\alpha$, $\beta$ and $\gamma$ are purely imaginary. In fact, we restrict to those cases that are relevant for the GZ framework, where the square masses are either $0$ or $\pm im^2$. More precisely, it can be shown that the relevant scalar sunset sum-integrals that appear in the GZ framework are $S_{\alpha\alpha\alpha}^{\kappa\lambda\tau}$, $S_{\alpha\alpha(-\alpha)}^{\kappa\lambda\tau}$, $S_{\alpha00}^{\kappa\lambda\tau}$, with $\smash{\alpha=\pm im^2}$, together of course with the corresponding permutations of masses \cite{vEgmond}.\\

Using the spectral representation (\ref{eq:spectral}) in Eq.~(\ref{eq:sunset}), we find
\beq
S^{\kappa\lambda\tau}_{\alpha\beta\gamma} & = & \int_{q_0;{\bf q}}\rho_\alpha(q_0;{\bf q})\int_{k_0;{\bf k}} \rho_\beta(k_0;{\bf k})\int_{l_0}\rho_\gamma(l_0;{\bf l})\,{\cal S}^{\kappa\lambda\tau}(q_0,k_0,l_0)\,,
\eeq
with
\beq\label{eq:90}
{\cal S}^{\kappa\lambda\tau}(q_0,k_0,l_0) & \equiv & T^2\sum_{n,m}\frac{1}{(q_0-i\omega_n^\kappa)(k_0-i\omega_m^\lambda)(l_0+i\omega_m^\lambda+i\omega_n^\kappa)}\,,
\eeq
a double Matsubara sum. Standard techniques for the evaluation of Matsubara sums together with color conservation $\kappa+\lambda+\tau=0$, lead then to
\beq
{\cal S}^{\kappa\lambda\tau}(q_0,k_0,l_0) & = & \frac{(n_{k_0-i\hat r\cdot\lambda}-n_{l_0+i\hat r\cdot\tau})(n_{q_0-i\hat r\cdot\kappa}-n_{-l_0-k_0-i\hat r\cdot\kappa})}{l_0+k_0+q_0}\nonumber\\
&  = & \frac{n_{k_0-i\hat r\cdot\lambda}n_{l_0-i\hat r\cdot\tau}+(-n_{-q_0+i\hat r\cdot\kappa})n_{k_0-i\hat r\cdot\lambda}+(-n_{-q_0+i\hat r\cdot\kappa})(-n_{-l_0+i\hat r\cdot\tau})}{l_0+k_0+q_0}\,,\label{eq:100}
\eeq
where, in going from the first to the second line, we have used the well known identity $n_x n_y=(1+n_x+n_y)n_{x+y}$. Finally, by making use of Eq.~(\ref{eq:11}), we arrive at the thermal splitting of the scalar sunset sum-integral: $S_{\alpha\beta\gamma}^{\kappa\lambda\tau}=S_{\alpha\beta\gamma}(0n)+S_{\alpha\beta\gamma}^{\kappa\lambda\tau}(1n)+S_{\alpha\beta\gamma}^{\kappa\lambda\tau}(2n)$. As in the previous example, $S_{\alpha\beta\gamma}(0n)$ denotes the pure vacuum contribution (no thermal factor), depending neither on the temperature nor on the background, while
\beq
S_{\alpha\beta\gamma}^{\kappa\lambda\tau}(1n)=\sum_{{\rm cyclic}}\int_{q_0;{\bf q}}\!\!\sigma_\alpha^\kappa(q_0;{\bf q})\,\int_{k_0;{\bf k}}\rho_\beta(k_0;{\bf k})\int_{l_0}\rho_\gamma(l_0;{\bf l})\,\frac{\theta(l_0)-\theta(-k_0)}{l_0+k_0+q_0}\,,
\eeq
and
\beq
S_{\alpha\beta\gamma}^{\kappa\lambda\tau}(2n)=\sum_{{\rm cyclic}}\int_{q_0;{\bf q}}\sigma_\alpha^\kappa(q_0;{\bf q})\int_{k_0;{\bf k}}\sigma_\beta^\lambda(k_0;{\bf k})\int_{l_0}\rho_\gamma(l_0;{\bf l})\,\frac{1}{l_0+k_0+q_0}\,,\label{eq:23}
\eeq
where $\sum_{{\rm cyclic}}$ stands for the cyclic permutations of the pairs $(\alpha,\kappa)$, $(\beta,\lambda)$ and $(\gamma,\tau)$ and $\sigma_\alpha^\kappa(q_0;{\bf q})$ was defined in the previous section.

We note that the thermal splitting considered here assumes that the denominator $l_0+k_0+q_0$ never vanishes. We show in Appendix \ref{app:regularization} that this is indeed so in those cases that are relevant for the GZ framework. More generic cases may require a regularization of the denominator but we shall not consider them here.

%%%
\subsection{Contribution with two thermal factors}
The contribution with two thermal factors is easily handled. We can first perform the $l_0$ integral by using the spectral representation (\ref{eq:spectral}) backwards. This leads to
\beq
S_{\alpha\beta\gamma}^{\kappa\lambda\tau}(2n)=\sum_{{\rm cyclic}}\int_{q_0;{\bf q}}\sigma_\alpha^\kappa(q_0;{\bf q})\int_{k_0;{\bf k}}\sigma_\beta^\lambda(k_0;{\bf k})\,\tilde G_\gamma(k_0+q_0;{\bf l})\,,
\eeq
where $\tilde G(z_0;{\bf z})$ is defined in Eq.~(\ref{eq:spectral}) and obeys $\tilde G(-z_0;{\bf z})=\tilde G(z_0;{\bf z})$. Next, we perform the $q_0$ and $k_0$ integrals and obtain
\beq\label{eq:B24}
S_{\alpha\beta\gamma}^{\kappa\lambda\tau}(2n)=\sum_{{\rm cyclic}}\sum_{\sigma_\alpha,\sigma_\beta}\int_{{\bf q}}\frac{n_{\varepsilon_{q,\alpha}-i\sigma_\alpha \hat r\cdot\kappa}}{2\varepsilon_{q,\alpha}}\int_{{\bf k}}\frac{n_{\varepsilon_{k,\beta}-i\sigma_\beta \hat r\cdot\lambda}}{2\varepsilon_{k,\beta}}\,\tilde G_\gamma(\sigma_\alpha\varepsilon_{q,\alpha}+\sigma_\beta\varepsilon_{k,\beta};{\bf l})\,,
\eeq
where $\sigma_\alpha$, $\sigma_\beta$ and $\sigma_\gamma$ take values in $\{-1,+1\}$. Finally, because this contribution is UV finite, we can set $d=4$ and perform the angular integrals. We find eventually
\beq
S_{\alpha\beta\gamma}^{\kappa\lambda\tau}(2n)=\frac{1}{64\pi^4}\sum_{{\rm cyclic}}\sum_{\sigma_\alpha,\sigma_\beta}\int_0^\infty dq\,q\frac{n_{\varepsilon_{q,\alpha}-i\sigma_\alpha \hat r\cdot\kappa}}{\varepsilon_{q,\alpha}}\int_0^\infty dk\,k\frac{n_{\varepsilon_{k,\beta}-i\sigma_\beta \hat r\cdot\lambda}}{\varepsilon_{k,\beta}}\ln\big(\alpha\beta;\gamma\big)\,,\label{eq:S2n}
\eeq
with
\beq
(\alpha\beta;\gamma\big)\equiv\frac{-(\sigma_\alpha\varepsilon_{q,\alpha}+\sigma_\beta\varepsilon_{k,\beta})^2+\varepsilon_{q+k,\gamma}^2}{-(\sigma_\alpha\varepsilon_{q,\alpha}+\sigma_\beta\varepsilon_{k,\beta})^2+\varepsilon_{q-k,\gamma}^2}=\frac{\gamma-\alpha-\beta-2\sigma_\alpha\sigma_\beta\varepsilon_{q,\alpha}\varepsilon_{k,\beta}+2qk}{\gamma-\alpha-\beta-2\sigma_\alpha\sigma_\beta\varepsilon_{q,\alpha}\varepsilon_{k,\beta}-2qk}\,.
\eeq

%%%
\subsection{Contribution with one thermal factor}
Integration over the frequencies leads this time to
\beq
S_{\alpha\beta\gamma}^{\kappa\lambda\tau}(1n)=\sum_{{\rm cyclic}}\sum_{\sigma_\alpha}\int_{\bf q}\frac{n_{\varepsilon_{q,\alpha}-i\sigma_\alpha \hat r\cdot\kappa}}{2\varepsilon_{q,\alpha}}\,\tilde I_{\beta\gamma}(\varepsilon_{q,\alpha};{\bf q})\,,
\eeq
with
\beq\label{eq:tildeI}
\tilde I_{\beta\gamma}(\varepsilon_{q,\alpha};{\bf q}) & =& \int_{{\bf k}}\left[\frac{1}{2\varepsilon_{l,\gamma}}\tilde G_\beta(\sigma_\alpha\varepsilon_{q,\alpha}+\varepsilon_{l,\gamma},{\bf k})+\frac{1}{2\varepsilon_{k,\beta}}\tilde G_\beta(\sigma_\alpha\varepsilon_{q,\alpha}-\varepsilon_{k,\beta},{\bf l})\right]\nonumber\\
& =& \int_{{\bf k}}\frac{1}{4\varepsilon_{k,\beta}\varepsilon_{l,\gamma}}\left[\frac{1}{\varepsilon_{k,\beta}+\varepsilon_{l,\gamma}+\sigma_\alpha\varepsilon_{q,\alpha}}+\frac{1}{\varepsilon_{k,\beta}+\varepsilon_{l,\gamma}-\sigma_\alpha\varepsilon_{q,\alpha}}\right]\nonumber\\
& = & \int_{{\bf k}} \frac{1}{2\varepsilon_{k,\beta}\varepsilon_{l,\gamma}}\frac{\varepsilon_{k,\beta}+\varepsilon_{l,\gamma}}{(\varepsilon_{k,\beta}+\varepsilon_{l,\gamma})^2-\varepsilon_{q,\alpha}^2}\,,
\eeq
where we note that the dependence on $\sigma_\alpha$ has dropped in the last line, which explains {\it a posteriori} why we did not include it in our notation for $\tilde I_{\beta\gamma}(\varepsilon_{q,\alpha};{\bf q})$. We show in Appendix \ref{app:I} that this quantity does not depend on ${\bf q}$ either. It follows that $S_{\alpha\beta\gamma}^{\kappa\lambda\tau}(1n)=\sum_{{\rm cyclic}} J_{\alpha}^\kappa(1n)I^\alpha_{\beta\gamma}(0n)$, with
\beq
I_{\beta\gamma}^\alpha(0n)\equiv\lim_{q\to 0}\tilde I_{\beta\gamma}(\varepsilon_{q,\alpha};{\bf q})=\int_{{\bf k}} \frac{1}{2\varepsilon_{k,\beta}\varepsilon_{k,\gamma}}\frac{\varepsilon_{k,\beta}+\varepsilon_{k,\gamma}}{(\varepsilon_{k,\beta}+\varepsilon_{k,\gamma})^2-\alpha}\,.
\eeq
The case $I_{\beta\gamma}^0(0n)$ can be evaluated immediately as
\beq\label{eq:remark}
I^0_{\beta\gamma}(0n) & = & \int_{k} \frac{1}{2\varepsilon_{k,\beta}\varepsilon_{k,\gamma}}\frac{1}{\varepsilon_{k,\beta}+\varepsilon_{k,\gamma}}\nonumber\\
& = & \int_{k} \frac{1}{2\varepsilon_{k,\beta}\varepsilon_{k,\gamma}}\frac{\varepsilon_{k,\beta}-\varepsilon_{k,\gamma}}{\beta-\gamma}=-\frac{J_\beta(0n)-J_\gamma(0n)}{\beta-\gamma}\,.
\eeq
As for the general case $I_{\beta\gamma}^\alpha(0n)$, we show in Appendix \ref{app:I} that it can be obtained from the analytic continuation of the vacuum Euclidean integral
\beq
I_{\beta\gamma}(0n)(Q^2\geq 0)\equiv \int_K^{T=0}G_\beta(K)G_\gamma(L)\,.
\eeq
We find
\beq\label{eq:I}
& & I_{\beta\gamma}^\alpha(0n)=\frac{1}{16\pi^2}\left\{\frac{1}{\epsilon}-\ln\frac{\bar\mu^2}{\alpha}+2\right.\nonumber\\
& & \hspace{3.2cm}+\,\frac{R(\alpha^-,\beta,\gamma)-\alpha-\beta+\gamma}{2\alpha}\ln\frac{R(\alpha^-,\beta,\gamma)-\alpha-\beta+\gamma}{2\bar\mu^2}\nonumber\\
& & \hspace{3.2cm}-\,\frac{R(\alpha^-,\beta,\gamma)+\alpha-\beta+\gamma}{2\alpha}\ln\frac{R(\alpha^-,\beta,\gamma)+\alpha-\beta+\gamma}{2\bar\mu^2}\nonumber\\
& & \hspace{3.2cm}-\,\frac{R(\alpha^-,\beta,\gamma)+\alpha+\beta-\gamma}{2\alpha}\ln\frac{R(\alpha^-,\beta,\gamma)+\alpha+\beta-\gamma}{2\bar\mu^2}\nonumber\\
& & \hspace{3.2cm}\left.+\,\frac{R(\alpha^-,\beta,\gamma)-\alpha+\beta-\gamma}{2\alpha}\ln\frac{R(\alpha^-,\beta,\gamma)-\alpha+\beta-\gamma}{2\bar\mu^2}\right\}\!,
\eeq
with $\alpha^-\equiv\alpha-0^+$ and
\beq
R^2(\alpha,\beta,\gamma)\equiv\alpha^2+\beta^2+\gamma^2-2\alpha\beta-2\beta\gamma-2\gamma\alpha\,.
\eeq
More details on the continuation are given in Appendix \ref{app:I}.

%%%
\subsection{Vacuum contribution (no thermal factor)}
The vacuum contribution can be written as a standard $d$-dimensional Euclidean integral
\beq
S_{\alpha\beta\gamma}(0n)  \equiv  \int_{Q,K}^{T=0} G_\alpha(Q)G_\beta(K)G_\gamma(L)\,.
\eeq
This integral is known analytically for positive square masses $\alpha$, $\beta$ and $\gamma$ \cite{Caffo:1998du}. One strategy to obtain the corresponding integral for imaginary square masses is to perform an analytic continuation of the result of \cite{Caffo:1998du} with respect to the masses. This is an efficient strategy in the case where the non-vanishing masses are all equal. 

Let us take for instance $S_{\alpha00}(0n)$. Seen as function of a complex $\alpha$, it is analytic with a branch cut for $\alpha \in \rm Re^-$. Because the expression for $\alpha \in \rm Re^+$,
\beq
S_{\alpha00}(0n)&=&(4\pi\mu^2)^{2\epsilon}\Gamma\left(1+\epsilon\right)^2\left(-\frac{\alpha}{128\pi^4}\right)\Bigg[ \frac{1}{4\epsilon^2} -\frac{1}{2\epsilon} \left(\ln \alpha-\frac{3}{2} \right)\nonumber\\
& & \hspace{5.0cm}+\,\frac{1}{2}  \Bigg(\!\ln^2 (\alpha)-3\ln (\alpha)+\frac{\pi^2}{6}+\frac{7}{2}\Bigg) \Bigg],\label{eq:S+00}
\eeq
has branch cuts only for $\alpha \in \rm Re^-$, it can be immediately used to represent $S_{\alpha00}(0n)$ in the case where $\alpha$ is purely imaginary. Using a similar argument, we obtain
\beq
S_{\alpha\alpha\alpha}(0n)&=&(4\pi\mu^2)^{2\epsilon}\Gamma\left(1+\epsilon\right)^2\left(-\frac{\alpha}{128\pi^4}\right)\Bigg[ \frac{3}{4\epsilon^2} -\frac{1}{2\epsilon} \left(3 \ln \alpha-\frac{9}{2} \right)\nonumber\\
&+&\left(\frac{3}{2} \ln ^2\left(\alpha\right)-\frac{9}{2} \ln \left(\alpha\right)- i \sqrt{3} \text{Li}_2\left(\frac{1}{2}
-\frac{i \sqrt{3}}{2}\right)+\frac{i \pi ^2}{12\sqrt{3}}+\frac{21}{4}\right) \Bigg].\label{eq:S+++}
\eeq

\vglue4mm 

The evaluation of $S_{\alpha\alpha(-\alpha)}(0n)$ is trickier because it involves the analytic continuation of a function of two complex variables. Although this can be done in principle, we here chose a more direct evaluation by adapting the technique in Ref.~\cite{Caffo:1998du} to the case of imaginary square masses. This technique is based on the derivation of a differential equation satisfied by $S_{\alpha\beta\gamma}(0n)$. Although the derivation of the differential equation is not affected by the presence of imaginary square masses, we reproduce it here for completeness and because we shall use it for other purposes later.

Let us write the scalar sunset vacuum integral symbolically as
\beq
\left\{G_\alpha G_\beta G_\gamma\right\}\equiv\int_{Q,K}^{T=0} G_\alpha(Q)G_\beta(K)G_\gamma(L)\,,
\eeq
with $L=-Q-K$. Expanding the identities
\beq
0=\left\{-\frac{\partial}{\partial Q_\mu}\Big(K_\mu\,G_\alpha G_\beta G_\gamma\Big)\right\}=\left\{-\frac{\partial}{\partial Q_\mu}\Big(Q_\mu\,G_\alpha G_\beta G_\gamma\Big)\right\}\!,\label{eq:div}
\eeq
we find
\beq
0 & = & \left\{2Q\cdot K\,G^2_\alpha G_\beta G_\gamma\right\}-\left\{2L\cdot K\,G_\alpha G_\beta G^2_\gamma\right\},\\
0 & = & -d\left\{G_\alpha G_\beta G_\gamma\right\}+\left\{2Q^2\,G^2_\alpha G_\beta G_\gamma\right\}-\left\{2Q\cdot L\,G_\alpha G_\beta G^2_\gamma\right\}.
\eeq
Then, writing $Q^2=Q^2+\alpha-\alpha$, as well as
\beq
2Q\cdot K & = & L^2+\gamma-Q^2-\alpha-K^2-\beta+(\alpha+\beta-\gamma)\,,\\
2L\cdot K & = & Q^2+\alpha-K^2-\beta-L^2-\gamma+(\beta+\gamma-\alpha)\,,\\
2Q\cdot L & = & K^2+\beta-L^2-\gamma-Q^2-\alpha+(\gamma+\alpha-\beta)\,,
\eeq
we arrive at
\beq
0 & = & \left\{G^2_\alpha G_\beta\right\}-\left\{G^2_\alpha G_\gamma\right\}+(\alpha+\beta-\gamma)\left\{G^2_\alpha G_\beta G_\gamma\right\}\nonumber\\
& - & \left\{G_\beta G^2_\gamma\right\}+\left\{G_\alpha G^2_\gamma\right\}-(\beta+\gamma-\alpha)\left\{G_\alpha G_\beta G^2_\gamma\right\},\\
0 & = & (3-d)\left\{G_\alpha G_\beta G_\gamma\right\}-2\alpha\left\{G^2_\alpha G_\beta G_\gamma\right\}\nonumber\\
& - & \left\{G_\alpha G^2_\gamma\right\}+\left\{G_\beta G^2_\gamma\right\}-(\gamma+\alpha-\beta)\left\{G_\alpha G_\beta G^2_\gamma\right\}.
\eeq
Using the first relation in order to replace $\{G^2_\alpha G_\beta G_\gamma\}$ in the second, we find
\beq
0 & = & (3-d)(\alpha+\beta-\gamma)\left\{G_\alpha G_\beta G_\gamma\right\}+R^2(\alpha,\beta,\gamma)\left\{G_\alpha G_\beta G^2_\gamma\right\}\nonumber\\
& + & (\alpha-\beta+\gamma)(\left\{G_\alpha G^2_\gamma\right\}-\left\{G_\beta G^2_\gamma\right\})+2\alpha (\left\{G^2_\alpha G_\beta\right\}-\left\{G^2_\alpha G_\gamma\right\})\,.
\eeq
Finally, using that $\{G_\alpha^2 G_\beta\}=\{G_\alpha^2\}\{G_\beta\}$, this rewrites
\beq
R^2(\alpha,\beta,\gamma) S_{\alpha\beta\gamma^2}(0n) & \,=\, & (d-3)(\alpha+\beta-\gamma)S_{\alpha\beta\gamma}(0n)+(d-2)J_\alpha(0n) J_\beta(0n)\nonumber\\
& - & (\alpha-\beta-\gamma)J_\alpha(0n) J_{\gamma^2}(0n)-(\beta-\alpha-\gamma)J_\beta(0n) J_{\gamma^2}(0n)\,,\label{eq:nicem1}
\eeq
where we have introduced the notations $J_{\alpha^2}(0n)\equiv \{G_\alpha^2\}$ and $S_{\alpha\beta\gamma^2}(0n)\equiv\left\{G_\alpha G_\beta G_{\gamma^2}\right\}$. In the case where $\gamma\neq 0$, we can use 
\beq
J_{\gamma^2}(0n) & = & -\frac{\partial}{\partial\gamma}J_\gamma(0n)=(1-d/2)\frac{J_\gamma(0n)}{\gamma}\,,\\
S_{\alpha\beta\gamma^2}(0n) & = & -\frac{\partial}{\partial\gamma}S_{\alpha\beta\gamma}(0n)\,,
\eeq
to arrive at the differential equation
\beq\label{eq:nice}
R^2(\alpha,\beta,\gamma)\frac{\partial}{\partial\gamma}S_{\alpha\beta\gamma}(0n) & \,=\, & (3-d)(\alpha+\beta-\gamma)S_{\alpha\beta\gamma}(0n)+(2-d)J_\alpha(0n) J_\beta(0n)\nonumber\\
& + & (2-d)\left[\frac{\alpha-\beta-\gamma}{2\gamma}J_\alpha(0n) J_\gamma(0n)+\frac{\beta-\alpha-\gamma}{2\gamma}J_\beta(0n) J_\gamma(0n)\right]\!.\nonumber\\
\eeq
In Appendix \ref{app:S++-}, we use this equation to obtain $S_{\alpha\alpha\gamma}(0n)$ in some appropriate range of values for $\gamma$. We follow the approach of Ref.~\cite{Caffo:1998du} by carefully adapting it to the case of imaginary square masses. We find in particular
\beq
& & S_{\alpha \alpha(-\alpha)}(0n)\nonumber\\
& & \hspace{0.5cm}=\,(4\pi\mu^2)^{2\epsilon}\Gamma\left(1+\epsilon\right)^2\left(-\frac{\alpha}{128\pi^4}\right)\Bigg[\frac{1}{4\epsilon^2}+\frac{1}{2\epsilon}\left(\frac{3}{2}-2 \ln \alpha+\ln (-\alpha)\right)\nonumber\\
& & \hspace{2.5cm}+\,\frac{\sqrt{5}}{2}\left(\frac{\pi ^2}{5}-i \pi\ln \frac{3-\sqrt{5}}{2}\right)\nonumber\\
&& \hspace{2.5cm}+\,\frac{7}{4}-3 \ln\alpha+\frac{3}{2}\ln(-\alpha)-\frac{1}{2}\ln\alpha \ln (-\alpha)+\frac{5}{4} \ln ^2 \alpha-\frac{1}{4} \ln ^2(-\alpha)\Bigg].\label{eq:S++-}
\eeq
We mention that the vacuum integrals $S_{+++}(0n)$, $S_{+00}(0n)$ and $S_{++-}(0n)$ have also been used in \cite{Gracey:2005cx,Gracey:2010cg} but the individual results are not quoted, only their final combination in the vacuum GZ horizon condition at two-loop order. For similar integrals involving both real and imaginary square masses, see \cite{Ford:2009ar}.

%%%
\subsection{Summary}
In summary, the scalar sunset sum-integral can be split as 
\beq
S_{\alpha\beta\gamma}^{\kappa\lambda\tau} & = & S_{\alpha\beta\gamma}(0n)+S_{\alpha\beta\gamma}^{\kappa\lambda\tau}(1n)+S_{\alpha\beta\gamma}^{\kappa\lambda\tau}(2n)\,,\label{eq:S012}
\eeq
with $S_{\alpha\beta\gamma}(0n)$ given in Eqs.~(\ref{eq:S+00}), (\ref{eq:S+++}) or (\ref{eq:S++-}) depending on the considered case, while $S_{\alpha\beta\gamma}^{\kappa\lambda\tau}(1n)$ reads
\beq
S_{\alpha\beta\gamma}^{\kappa\lambda\tau}(1n)=J_{\alpha}^\kappa(1n)I^\alpha_{\beta\gamma}(0n)+J_{\beta}^\lambda(1n)I^\beta_{\gamma\alpha}(0n)+J_{\gamma}^\tau(1n)I^\gamma_{\alpha\beta}(0n)\,,
\eeq 
with $I_{\beta\gamma}^\alpha(0n)$ given in Eq.~(\ref{eq:I}), and $S_{\alpha\beta\gamma}^{\kappa\lambda\tau}(2n)$ is given in Eq.~(\ref{eq:S2n}).

%%%%%
\section{Mass derivatives}\label{sec:mass_derivatives}
The various sunset diagrams that appear in the GZ framework lead also to mass derivatives of the scalar sunset, in the limit where the corresponding mass is taken to zero. In fact, what appear are the limits \cite{vEgmond}
\beq
\Delta S^{\kappa\lambda\tau}_{0^2\beta\gamma} & \equiv & \lim_{\alpha\to 0}\left[S^{\kappa\lambda\tau}_{\alpha^2\beta\gamma}+J_{\alpha^2}^\kappa \frac{J_\beta^\lambda-J_\gamma^\tau}{\beta-\gamma}\right]\!,\label{eq:D1}\\
\Delta S^{\kappa\lambda\tau}_{0^20^2\gamma} & \equiv & \lim_{\alpha\to 0}\lim_{\beta\to 0}\left[S^{\kappa\lambda\tau}_{\alpha^2\beta^2\gamma}-\frac{J_{\alpha^2}^\kappa J_{\beta^2}^\lambda}{\gamma}-\frac{J_\gamma^\tau-J_0^\kappa}{\gamma^2} J_{\beta^2}^\lambda- \frac{J_\gamma^\tau-J_0^\lambda}{\gamma^2}J_{\alpha^2}^\kappa\right]\!,\label{eq:D2}
\eeq
where, as already introduced above, the squaring of the mass indices corresponds to the doubling of the associated propagators, or more generally to taking minus the derivative with respect to the associated square mass. The relevant cases for the GZ framework are $\Delta S^{\kappa\lambda\tau}_{0^2\gamma(-\gamma)}$, $\Delta S^{\kappa\lambda\tau}_{0^20\gamma}$, $\Delta S^{\kappa\lambda\tau}_{0^20^2\gamma}$, with $\smash{\gamma=\pm im^2}$, together with the corresponding permutations of the masses \cite{vEgmond}. We also mention that $\Delta S_{0^2,\alpha\beta}$ is nothing but the function $\bar T$ defined in \cite{Martin:2003qz}, for zero external momentum but generalized to the case of finite temperature.

It is easily seen that, even though $S^{\kappa\lambda\tau}_{\alpha^2\beta\gamma}$, $S^{\kappa\lambda\tau}_{\alpha^2\beta^2\gamma}$, $J_{\alpha^2}^\kappa$ and $J_{\beta^2}^\lambda$ are singular in the limits $\alpha\to 0$ or $\beta\to 0$, the combinations of sum-integrals in Eqs.~(\ref{eq:D1}) and (\ref{eq:D2}) admit regular limits. To verify this, let us first note that the potential singularities originate either from the vacuum pieces which do not depend on the color weights ($\kappa$, $\lambda$, $\tau$), or from the thermal pieces in the case where the weights are equal to zero. Therefore, one can safely ignore the weights in order to check the regularity of the above limits. For instance, in the limit $\alpha\to 0$, the sum-integral $S_{\alpha^2\beta\gamma}$ is dominated by the $Q\to 0$ region and behaves consequently as
\beq
S_{\alpha^2\beta\gamma}\sim \int_Q\frac{1}{(Q^2+\alpha)^2}\times \int_K \frac{1}{(K^2+\beta)(K^2+\gamma)}=-J_{\alpha^2}\frac{J_\beta-J_\gamma}{\beta-\gamma}\,.
\eeq
The divergent behavior in the RHS is precisely what is subtracted in Eq.~(\ref{eq:D1}) to ensure that the limit is regular.\footnote{Because the divergence is at most $\sim\alpha^{-1/2}$ at finite temperature, there are no subleading divergent terms.} To understand the singular structure of $S_{\alpha^2\beta^2\gamma}$, we first write it identically as
\beq
S_{\alpha^2\beta^2\gamma} & = & \frac{1}{\gamma}\int_{Q,K}\frac{1}{(Q^2+\alpha)^2}\frac{1}{(K^2+\beta)^2}\nonumber\\
& - & \frac{1}{\gamma}\int_{Q,K}\frac{1}{(Q^2+\alpha)^2}\frac{1}{(K^2+\beta)^2}\frac{K^2+Q^2+2K\cdot Q}{(K+Q)^2+\gamma}\,.
\eeq
The term with $K\cdot Q$ in the second line leads to regular contributions in the limit $\alpha\to 0$ and $\beta\to 0$, while the term with $K^2$ (resp. $Q^2$) leads to singular contributions as $\alpha\to 0$ (resp. $\beta\to 0$), controlled by the $Q\to0$ (resp. $K\to 0$) region of the integral. We find
\beq
S_{\alpha^2\beta^2\gamma} & = & \frac{1}{\gamma}\int_{Q}\frac{1}{(Q^2+\alpha)^2}\int_K\frac{1}{(K^2+\beta)^2}\nonumber\\
& - & \frac{1}{\gamma}\int_{Q}\frac{1}{(Q^2+\alpha)^2}\int_K\frac{1}{K^2(K^2+\gamma)}\nonumber\\
& - & \frac{1}{\gamma}\int_{Q}\frac{1}{Q^2(Q^2+\alpha)}\int_K\frac{1}{(K^2+\beta)^2}+{\rm regular}\nonumber\\
& = & \frac{J_{\alpha^2}J_{\beta^2}}{\gamma}+\frac{J_\gamma-J_0}{\gamma^2}J_{\alpha^2}+\frac{J_\gamma-J_0}{\gamma^2}J_{\beta^2}+{\rm regular}\,.
\eeq
These are precisely the terms that are subtracted in Eq.~(\ref{eq:D2}) to obtain a regular limit.

The regular limits $\Delta S^{\kappa\lambda\tau}_{0^2\beta\gamma}$ and $\Delta S^{\kappa\lambda\tau}_{0^20^2\gamma}$ admit thermal splittings that one derives from the corresponding splitting (\ref{eq:S012}) of the scalar sunset, and which we now discuss. 

%%%
\subsection{Thermal splitting}
From Eqs.~(\ref{eq:S012}) and (\ref{eq:D1}), we find the thermal splitting
\beq
\Delta S^{\kappa\lambda\tau}_{0^2\beta\gamma}=\Delta S_{0^2\beta\gamma}(0n)+\Delta S^{\kappa\lambda\tau}_{0^2\beta\gamma}(1n)+\Delta S^{\kappa\lambda\tau}_{0^2\beta\gamma}(2n)\,,
\eeq
with
\beq
\Delta S_{0^2\beta\gamma}(0n) & \!=\! &  \lim_{\alpha\to 0}\!\left[S_{\alpha^2\beta\gamma}(0n)\!+\!J_{\alpha^2}(0n) \frac{J_\beta(0n)-J_\gamma(0n)}{\beta-\gamma}\right]\!,\label{eq:DS0}\\
\Delta S^{\kappa\lambda\tau}_{0^2\beta\gamma}(1n) & \!=\! &  \lim_{\alpha\to 0}\!\left[S^{\kappa\lambda\tau}_{\alpha^2\beta\gamma}(1n)\!+\!J_{\alpha^2}(0n) \frac{J_\beta^\lambda(1n)-J_\gamma^\tau(1n)}{\beta-\gamma}\!+\!J_{\alpha^2}^\kappa(1n) \frac{J_\beta(0n)-J_\gamma(0n)}{\beta-\gamma}\right]\!,\label{eq:DS1}\\
\Delta S^{\kappa\lambda\tau}_{0^2\beta\gamma}(2n) & \!=\! & \lim_{\alpha\to 0}\!\left[S^{\kappa\lambda\tau}_{\alpha^2\beta\gamma}(2n)\!+\!J_{\alpha^2}^\kappa(1n) \frac{J_\beta^\lambda(1n)-J_\gamma^\tau(1n)}{\beta-\gamma}\right]\!.
\eeq
Similarly, from Eqs.~(\ref{eq:S012}) and (\ref{eq:D2}), we find the thermal splitting
\beq
\Delta S^{\kappa\lambda\tau}_{0^20^2\gamma}=\Delta S_{0^20^2\gamma}(0n)+\Delta S^{\kappa\lambda\tau}_{0^20^2\gamma}(1n)+\Delta S^{\kappa\lambda\tau}_{0^20^2\gamma}(2n)\,,
\eeq 
with
\beq
\Delta S_{0^20^2\gamma}(0n) & \!=\! & \lim_{\alpha\to 0}\lim_{\beta\to 0}\left[S_{\alpha^2\beta^2\gamma}(0n)-\frac{J_{\alpha^2}(0n)J_{\beta^2}(0n)}{\gamma}-\frac{J_{\alpha^2}(0n)+J_{\beta^2}(0n)}{\gamma^2} J_\gamma(0n)\right]\!,\label{eq:eq}\\
\Delta S^{\kappa\lambda\tau}_{0^20^2\gamma}(1n) & \!=\! & \lim_{\alpha\to 0}\lim_{\beta\to 0}\left[S^{\kappa\lambda\tau}_{\alpha^2\beta^2\gamma}(1n)-\frac{J_{\alpha^2}(0n) J_{\beta^2}^\lambda(1n)}{\gamma}-\frac{J_{\alpha^2}^\kappa(1n) J_{\beta^2}(0n)}{\gamma}\right.\nonumber\\
&  & \hspace{3.6cm}-\,\frac{J_\gamma(0n)}{\gamma^2} J_{\beta^2}^\lambda(1n)-\frac{J_\gamma^\tau(1n)-J_0^\kappa(1n)}{\gamma^2} J_{\beta^2}(0n)\nonumber\\
&  & \hspace{3.6cm}\left.-\,\frac{J_\gamma(0n)}{\gamma^2}J_{\alpha^2}^\kappa(1n)-\frac{J_\gamma^\tau(1n)-J_0^\lambda(1n)}{\gamma^2}J_{\alpha^2}(0n)\right]\!,\\
\Delta S^{\kappa\lambda\tau}_{0^20^2\gamma}(2n) & \!=\! & \lim_{\alpha\to 0}\lim_{\beta\to 0}\left[S^{\kappa\lambda\tau}_{\alpha^2\beta^2\gamma}(2n)-\frac{J_{\alpha^2}^\kappa(1n)J_{\beta^2}^\lambda(1n)}{\gamma}\right.\nonumber\\
& & \hspace{1.7cm}\left.-\,\frac{J_\gamma^\tau(1n)-J_0^\kappa(1n)}{\gamma^2} J_{\beta^2}^\lambda(1n)-\frac{J_\gamma^\tau(1n)-J_0^\lambda(1n)}{\gamma^2}J_{\alpha^2}^\kappa(1n)\right]\!,
\eeq
where we have used that $\smash{J_0(0n)=0}$. In what follows, we shall also make extensive use of the property $\smash{J_{0^2}(0n)=0}$, valid in dimensional regularization \cite{}. This property might be more difficult to grasp than the previous one because $J_{\alpha^2}(0n)$ diverges in the limit $\smash{\alpha\to 0}$. However this just means that the function $J_{\alpha^2}(0n)$, although defined for $\smash{\alpha=0}$, is not continuous at $\smash{\alpha=0}$. Then, we shall always make sure that when the property $\smash{J_{0^2}(0n)=0}$ is used, it corresponds to $J_{\alpha^2}(0n)$ being evaluated for $\smash{\alpha=0}$ and not to a limit being taken. We mention finally that the results to be presented below can be obtained without ever using $\smash{J_{0^2}(0n)=0}$ although the calculations are lengthier.

%%%
\subsection{Vacuum contributions (no thermal factor)}
\underline{$\Delta S_{0^2\beta\gamma}(0n)$:} The vacuum contribution $\Delta S_{0^2\beta\gamma}(0n)$ is in fact nothing but $S_{0^2\beta\gamma}(0n)$. Indeed, even though both $S_{\alpha^2\beta\gamma}(0n)$ and $J_{\alpha^2}(0n)$ are singular in the limit $\smash{\alpha\to 0}$, their particular combination in Eq.~(\ref{eq:DS0}) is regular. Moreover since $S_{0^2\beta\gamma}(0n)$ and $J_{0^2}(0n)$ are well defined in dimensional regularization, the limit $\smash{\gamma\to 0}$ is then equivalent to the direct evaluation at $\smash{\gamma=0}$, and we find $\smash{\Delta S_{0^2\beta\gamma}(0n)=S_{0^2\beta\gamma}(0n)}$ owing to the fact that $\smash{J_{0^2}(0n)=0}$. Now, since Eq.~(\ref{eq:nicem1}) is valid for $\smash{\gamma=0}$, and using once more the property $\smash{J_{0^2}(0n)=0}$, we arrive at
\beq
S_{0^2\beta\gamma}(0n)=(d-3)\frac{\beta+\gamma}{(\beta-\gamma)^2}S_{0\beta\gamma}(0n)+(d-2)\frac{J_\beta(0n) J_\gamma(0n)}{(\beta-\gamma)^2}\,,\label{eq:vac_result}
\eeq
which expresses $\Delta S_{\alpha\beta0^2}(0n)$ in terms of already determined functions. We notice that, in the case of $\Delta S_{0^2\gamma(-\gamma)}$, the term proportional to $S_{0\gamma(-\gamma)}(0n)$ vanishes and therefore we do not need to consider this vacuum sunset integral.\\

\underline{$\Delta S_{0^20^2\gamma}(0n)$:} We can proceed similarly for $\Delta S_{0^20^2\gamma}(0n)$. First, from the same argument as above, we find $\smash{\Delta S_{0^20^2\gamma}(0n)=S_{0^20^2\gamma}(0n)}$. The difference with the above is that we do not have an equation fixing directly $S_{0^20^2\gamma}(0n)$.  Acting on Eq.~(\ref{eq:vac_result}) with $-\partial/\partial\beta$, we obtain an equation for $S_{0^2\beta^2\gamma}(0n)$ with $\smash{\beta\neq 0}$, but $S_{0^20^2\gamma}(0n)$ is not the limit of $S_{0^2\beta^2\gamma}(0n)$. The way out is to subtract from $S_{0^2\beta^2\gamma}$ its $\smash{\beta\to 0}$ divergent part, in such a way that
 \beq
 S_{0^20^2\gamma}(0n)=\lim_{\beta\to 0}\left[S_{0^2\beta^2\gamma}(0n)-\frac{J_{\beta^2}(0n)J_\gamma(0n)}{\gamma^2}\right],
 \eeq
 owing again to $\smash{J_{0^2}(0n)=0}$. We find
\beq
& & S_{0^2\beta^2\gamma}(0n)-\frac{J_{\beta^2}(0n)J_\gamma(0n)}{\gamma^2}\nonumber\\
& & \hspace{0.5cm}=\,(d-3)\frac{\beta+3\gamma}{(\beta-\gamma)^3}S_{0\beta\gamma}(0n)+(d-3)\frac{\beta+\gamma}{(\beta-\gamma)^2}\left[S_{0\beta^2\gamma}(0n)+\frac{J_{\beta^2}(0n)J_\gamma(0n)}{\gamma}\right]\nonumber\\
& & \hspace{0.5cm}+\,2(d-2)\frac{J_\beta(0n)J_\gamma(0n)}{(\beta-\gamma)^3}+\left[\frac{d-2}{(\beta-\gamma)^2}-\frac{1}{\gamma^2}
-\frac{d-3}{(\beta-\gamma)^2}\frac{\beta+\gamma}{\gamma}\right]J_{\beta^2}(0n)J_\gamma(0n)\,.
\eeq
The dangerous contributions proportional to $J_{\beta^2}(0n)$ in the RHS cancel in the limit $\smash{\beta\to 0}$ and we find eventually
 \beq
S_{0^20^2\gamma}(0n) & = & -3(d-3)\frac{S_{00\gamma}(0n)}{\gamma^2}+(d-3)\frac{S_{00^2\gamma}(0n)}{\gamma}\,,\nonumber\\
  & = &  (d-3)(d-6)\frac{S_{00\gamma}(0n)}{\gamma^2}\,,\label{eq:idd}
 \eeq
 where we have once more made use of Eq.~(\ref{eq:vac_result}).  We have cross-checked this last result using a direct evaluation of $S_{0^20^2\gamma}(0n)$ using standard techniques.

%%%
\subsection{Contributions with one thermal factor}
\underline{$\Delta S^{\kappa\lambda\tau}_{0^2\beta\gamma}(1n)$:} The contribution with one thermal factor to $\Delta S^{\kappa\lambda\tau}_{0^2\beta\gamma}$ can be rewritten as
\beq
\Delta S^{\kappa\lambda\tau}_{0^2\beta\gamma}(1n) & \!=\! & \lim_{\alpha\to 0}\Bigg[J_{\alpha^2}^\kappa(1n)I^\alpha_{\beta\gamma}(0n)+J_{\alpha}^\kappa(1n)I^{\alpha^2}_{\beta\gamma}(0n)+J_{\beta}^\lambda(1n)I^\beta_{\alpha^2\gamma}(0n)+J_{\gamma}^\tau(1n)I^\gamma_{\alpha^2\beta}(0n)\nonumber\\
& & \hspace{0.8cm}+\,J_{\alpha^2}(0n) \frac{J_\beta^\lambda(1n)-J_\gamma^\tau(1n)}{\beta-\gamma}+J_{\alpha^2}^\kappa(1n) \frac{J_\beta(0n)-J_\gamma(0n)}{\beta-\gamma}\Bigg]\nonumber\\
& \!=\! & \lim_{\alpha\to 0}\Bigg[J_{\alpha^2}^\kappa(1n)\left(I^\alpha_{\beta\gamma}(0n)+\frac{J_\beta(0n)-J_\gamma(0n)}{\beta-\gamma}\right)+J_{\alpha}^\kappa(1n)I^{\alpha^2}_{\beta\gamma}(0n)\nonumber\\
& & \hspace{0.8cm}+\,J_\beta^\lambda(1n)\left(I^\beta_{\alpha^2\gamma}(0n)+\frac{J_{\alpha^2}(0n)}{\beta-\gamma}\right)\!+\!J_\gamma^\tau(1n)\left(I^\gamma_{\alpha^2\beta}(0n)+\frac{J_{\alpha^2}(0n)}{\gamma-\beta}\right)\Bigg].
\eeq
Owing to Eq.~(\ref{eq:remark}), it seems that the contribution in the first round bracket can be neglected in the limit $\smash{\alpha\to 0}$. This turns out to be true, although one needs to pay a little bit of attention, since, at finite temperature and in the case where $\smash{\kappa=0}$, $J_{\alpha^2}^\kappa(1n)$  diverges in the same limit. Fortunately, the divergence goes as $\alpha^{-1/2}$, which is not enough to compensate the vanishing of the round bracket $\sim\alpha$. We find eventually that
\beq\label{eq:DS12}
\Delta S^{\kappa\lambda\tau}_{0^2\beta\gamma}(1n)=J_{0}^\kappa(1n)\Delta I^{0^2}_{\beta\gamma}(0n)+J_\beta^\lambda(1n)\Delta I^\beta_{0^2\gamma}(0n)+J_\gamma^\tau(1n)\Delta I^\gamma_{0^2\beta}(0n)\,,
\eeq
with
\beq
\Delta I^{0^2}_{\beta\gamma}(0n) & \equiv & I^{0^2}_{\beta\gamma}(0n)\,,\\
\Delta I^\beta_{0^2\gamma}(0n) & \equiv & \lim_{\alpha\to 0}\left[I^\beta_{\alpha^2\gamma}(0n)+\frac{J_{\alpha^2}(0n)}{\beta-\gamma}\right]=I^\beta_{0^2\gamma}(0n)\,,
\eeq
where the contribution within brackets is regular in the limit $\smash{\alpha\to 0}$ and we have used $\smash{J_{0^2}(0n)=0}$ in the last step.

The first quantity can be computed using similar tricks as for $I_{\beta\gamma}^0(0n)$ in the previous section, namely
\beq
\Delta I^{0^2}_{\beta\gamma}(0n) & = & -\int_{k} \frac{1}{2\varepsilon_{k,\beta}\varepsilon_{k,\gamma}}\frac{1}{(\varepsilon_{k,\beta}+\varepsilon_{k,\gamma})^3}=-\int_{k} \frac{1}{2\varepsilon_{k,\beta}\varepsilon_{k,\gamma}}\frac{(\varepsilon_{k,\beta}-\varepsilon_{k,\gamma})^3}{(\beta-\gamma)^3}\nonumber\\
& = & -\int_{k} \frac{(\varepsilon^2_{k,\beta}+3\varepsilon^2_{k,\gamma})/\varepsilon_{k,\gamma}-(\varepsilon^2_{k,\gamma}+3\varepsilon^2_{k,\beta})/\varepsilon_{k,\beta}}{2(\beta-\gamma)^3}\nonumber\\
& = & \frac{1}{(\gamma-\beta)^3}\left[\left(\beta+\frac{4-d}{d}\gamma\right)J_\gamma(0n)-\left(\gamma+\frac{4-d}{d}\beta\right)J_\beta(0n)\right],\label{eq:eqq}
\eeq
where we have used that
\beq
\int_k \frac{k^2}{2\varepsilon_{k,\beta}}=\int_K^{T=0}\frac{k^2}{K^2+\beta}=-\frac{d-1}{d}\beta J_\beta(0n)\,,
\eeq
together with $\int_K^{T=0}1=0$ in dimensional regularization. 

As for $\Delta I^\beta_{0^2\gamma}(0n)$, we can proceed in many different ways, either by acting with $-\partial/\partial\alpha$ on the previously determined expression for $I_{\alpha\gamma}^\beta(0n)$, followed by the $\alpha\to 0$ limit after appropriate subtraction of the $\smash{\alpha\to 0}$ singular part, or by computing the appropriate subtracted Euclidean integral 
\beq
\Delta I_{0^2\gamma}(0n)(K^2)\equiv \lim_{\alpha\to 0}\left[I_{\alpha^2\gamma}(K^2)-\frac{J_{\alpha^2}(0n)}{K^2+\gamma}\right]=I_{0^2\gamma}(K^2)\,,\label{eq:84}
\eeq
and analytically continuing it from $K^2>0$ to $K^2=-\beta$ imaginary. Here we proceed with this second strategy but instead of continuing the explicit expression of the integral, we continue the corresponding differential equation, with the advantage that $\Delta I^\beta_{0^2\gamma}(0n)$ will be expressed in terms of already computed quantities. 

To derive the differential equation, we basically consider the same equations as in Eqs.~(\ref{eq:div}) but with the propagator $G_\beta(K)$ and the integral $\int_K^{T=0}$ missing. It is easily seen that one can follow the steps below Eqs.~(\ref{eq:div}) by removing the factors $G_\beta(K)$ (those terms that did not have such a factor need to be discarded) and to replace the explicit occurrences of $\beta$ by $-K^2$. It follows that
\beq
R^2(\gamma,-K^2,\alpha) I_{\alpha^2\gamma}(0n)(K^2) & = & (d-3)(\gamma-K^2-\alpha)I_{\alpha\gamma}(0n)(K^2)\nonumber\\
& + & (d-2)J_\gamma(0n)+(K^2+\alpha+\gamma) J_{\alpha^2}(0n)\,.
\eeq
This identity is valid for $\smash{\alpha=0}$, in which case, we obtain
\beq
I_{0^2\gamma}(0n)(K^2)=\frac{(d-3)(\gamma-K^2)I_{0\gamma}(0n)(K^2)+(d-2)J_\gamma(0n)}{(K^2+\gamma)^2}\,.\label{eq:86}
\eeq
After continuation, we find eventually
\beq
I^\beta_{0^2\gamma}(0n)=\frac{(d-3)(\beta+\gamma)I^\beta_{0\gamma}(0n)+(d-2)J_\gamma(0n)}{(\beta-\gamma)^2}\,.\label{eq:final2}
\eeq

\underline{$\Delta S^{\kappa\lambda\tau}_{0^20^2\gamma}(1n)$:} Similarly, the contribution with one thermal factor to $\Delta S^{\kappa\lambda\tau}_{0^20^2\gamma}$ can be rewritten as
\beq
\Delta S^{\kappa\lambda\tau}_{0^20^2\gamma}(1n) & = & \lim_{\alpha\to 0}\lim_{\beta\to 0}\Bigg[J_{\alpha^2}^\kappa(1n)\left(I^\alpha_{\beta^2\gamma}(0n)-\frac{J_{\beta^2}(0n)}{\gamma}- \frac{J_\gamma(0n)}{\gamma^2}\right)\nonumber\\
& & \hspace{1.6cm}+\,J_{\beta^2}^\lambda(1n))\left(I^\beta_{\alpha^2\gamma}(0n)-\frac{J_{\alpha^2}(0n)}{\gamma}- \frac{J_\gamma(0n)}{\gamma^2}\right)\nonumber\\
& & \hspace{1.6cm}+\,J_{\alpha}^\kappa(1n)\left(I^{\alpha^2}_{\beta^2\gamma}(0n)+\frac{J_{\beta^2}(0n)}{\gamma^2}\right)\nonumber\\
& & \hspace{1.6cm}+\,J_{\beta}^\lambda(1n)\left(I^{\beta^2}_{\alpha^2\gamma}(0n)+\frac{J_{\alpha^2}(0n)}{\gamma^2}\right)\nonumber\\
& & \hspace{1.6cm}+\,J_{\gamma}^\tau(1n)\left(I^\gamma_{\alpha^2\beta^2}(0n)-\frac{J_{\alpha^2}(0n)}{\gamma^2}-\frac{J_{\beta^2}(0n)}{\gamma^2}\right)\Bigg].
\eeq
Using Eq.~(\ref{eq:remark}), this rewrites
\beq
\Delta S^{\kappa\lambda\tau}_{0^20^2\gamma}(1n)= \big(J_{0}^\kappa(1n)+J_{0}^\lambda(1n)\big)\Delta I^{0^2}_{0^2\gamma}(0n)+J_{\gamma}^\tau(1n)\Delta I^\gamma_{0^20^2}(0n)\,,
\eeq
with
\beq
\Delta I^{0^2}_{0^2\gamma}(0n) & \equiv & \lim_{\alpha\to 0}\lim_{\beta\to 0}\Bigg[I^{\alpha^2}_{\beta^2\gamma}(0n)+\frac{J_{\beta^2}(0n)}{\gamma^2}\Bigg]=I^{0^2}_{0^2\gamma}(0n),\\
\Delta I^\gamma_{0^20^2}(0n) & \equiv & \lim_{\alpha\to 0}\lim_{\beta\to 0}\Bigg[I^\gamma_{\alpha^2\beta^2}(0n)-\frac{J_{\alpha^2}(0n)}{\gamma^2}-\frac{J_{\beta^2}(0n)}{\gamma^2}\Bigg]=I^\gamma_{0^20^2}(0n).\label{eq:91}
\eeq
We note that
\beq
I^{0^2}_{0^2\gamma}(0n) & = & \lim_{\beta\to 0}\Bigg[I^{0^2}_{\beta^2\gamma}(0n)+\frac{J_{\beta^2}(0n)}{\gamma^2}\Bigg],\label{eq:sub}\\
I^\gamma_{0^20^2}(0n) & = & \lim_{\beta\to 0}\Bigg[I^\gamma_{0^2\beta^2}(0n)-\frac{J_{\beta^2}(0n)}{\gamma^2}\Bigg].\label{eq:sub2}
\eeq
Using Eqs.~(\ref{eq:eqq}) and (\ref{eq:final2}) and after subtracting the $\smash{\beta\to 0}$ and $\smash{\gamma\to 0}$ singular parts, according to Eqs.~(\ref{eq:sub}) and (\ref{eq:sub2}) respectively, we find eventually
\beq
I_{0^2\gamma}^{0^2}(0n) & = & \frac{d-6}{d/2}\frac{J_\gamma(0n)}{\gamma^3}\,,\\
I^\beta_{0^20^2}(0n) & = & (d-3)(d-6)\frac{I^\beta_{00}(0n)}{\beta^2}\,.\label{eq:98}
\eeq

%%%
\subsection{Contributions with two thermal factors}
\underline{$\Delta S^{\kappa\lambda\tau}_{0^2\beta\gamma}(2n)$:} The contribution with two thermal factors to $\Delta S^{\kappa\lambda\tau}_{0^2\beta\gamma}$ can be rewritten as
\beq
\Delta S_{0^2\beta\gamma}^{\kappa\lambda\tau}(2n) &=& -\frac{1}{64\pi^4}\lim_{\alpha\to 0}\Bigg\{\nonumber\\
&& +\,\sum_{\sigma_\alpha,\sigma_\beta}\int_0^\infty dq\,q\,\frac{d}{d\alpha}\left(\frac{n_{\varepsilon_{q,\alpha}-i\sigma_\alpha \hat r\cdot\kappa}}{\varepsilon_{\alpha,q}}\right)\int_0^\infty dk\,k\,\frac{n_{\varepsilon_{k,\beta}-i\sigma_\beta \hat r\cdot\lambda}}{\varepsilon_{\beta,k}}\left[\ln\big(\alpha\beta;\gamma\big)+\frac{4qk}{\beta-\gamma}\right]\nonumber\\
&& +\,\sum_{\sigma_\alpha,\sigma_\gamma}\int_0^\infty dq\,q\,\frac{d}{d\alpha}\left(\frac{n_{\varepsilon_{q,\alpha}-i\sigma_\alpha \hat r\cdot\kappa}}{\varepsilon_{\alpha,q}}\right)\int_0^\infty dk\,k\,\frac{n_{\varepsilon_{k,\gamma}-i\sigma_\gamma \hat r\cdot\tau}}{\varepsilon_{\gamma,k}}\left[\ln\big(\alpha\gamma;\beta\big)+\frac{4qk}{\gamma-\beta}\right]\nonumber\\
&& +\,\sum_{\sigma_\alpha,\sigma_\beta}\int_0^\infty dq\,q\,\frac{n_{\varepsilon_{q,\alpha}-i\sigma_\alpha \hat r\cdot\kappa}}{\varepsilon_{\alpha,q}}\int_0^\infty dk\,k\,\frac{n_{\varepsilon_{k,\beta}-i\sigma_\beta \hat r\cdot\lambda}}{\varepsilon_{\beta,k}}\,\frac{d}{d\alpha}\ln\big(\alpha\beta;\gamma\big)\nonumber\\
&& +\,\sum_{\sigma_\alpha,\sigma_\gamma}\int_0^\infty dq\,q\,\frac{n_{\varepsilon_{q,\alpha}-i\sigma_\alpha \hat r\cdot\kappa}}{\varepsilon_{\alpha,q}}\int_0^\infty dk\,k\,\frac{n_{\varepsilon_{k,\gamma}-i\sigma_\gamma \hat r\cdot\tau}}{\varepsilon_{\gamma,k}}\,\frac{d}{d\alpha}\ln\big(\alpha\gamma;\beta\big)\nonumber\\
&& +\,\sum_{\sigma_\beta,\sigma_\gamma}\int_0^\infty dq\,q\,\frac{n_{\varepsilon_{q,\beta}-i\sigma_\beta \hat r\cdot\lambda}}{\varepsilon_{\beta,q}}\int_0^\infty dk\,k\,\frac{n_{\varepsilon_{k,\gamma}-i\sigma_\gamma \hat r\cdot\tau}}{\varepsilon_{\gamma,k}}\,\frac{d}{d\alpha}\ln\big(\beta\gamma;\alpha\big)\Bigg\}.
\eeq
The terms with the $\alpha$-derivative acting on the thermal factor can be treated using an integration by parts after noticing that $df(\varepsilon_{q,\alpha})/d\alpha= df(\varepsilon_{q,\alpha})/dq^2=(df(\varepsilon_{q,\alpha})/dq)/(2q)$. The boundary term vanishes both for $q\to\infty$ (due to the thermal factor). The boundary at $q=0$ contributes
\beq
& & \frac{1}{128\pi^4}\lim_{\alpha\to 0}\Bigg\{\sum_{\sigma_\alpha,\sigma_\beta}\frac{n_{\varepsilon_{q,\alpha}-i\sigma_\alpha \hat r\cdot\kappa}}{\varepsilon_{\alpha,q}}\int_0^\infty dk\,k\,\frac{n_{\varepsilon_{k,\beta}-i\sigma_\beta \hat r\cdot\lambda}}{\varepsilon_{\beta,k}}\lim_{q\to 0}\left[\ln\big(\alpha\beta;\gamma\big)+\frac{4qk}{\beta-\gamma}\right]\nonumber\\
&& +\,\sum_{\sigma_\alpha,\sigma_\gamma}\frac{n_{\varepsilon_{q,\alpha}-i\sigma_\alpha \hat r\cdot\kappa}}{\varepsilon_{\alpha,q}}\int_0^\infty dk\,k\,\frac{n_{\varepsilon_{k,\gamma}-i\sigma_\gamma \hat r\cdot\tau}}{\varepsilon_{\gamma,k}}\lim_{q\to 0}\left[\ln\big(\alpha\gamma;\beta\big)+\frac{4qk}{\gamma-\beta}\right]\Bigg\}=0\,.
\eeq 
(note that $\alpha\to 0$ is taken only after the limit $q\to 0$). We obtain
\beq
\Delta S_{0^2\beta\gamma}^{\kappa\lambda\tau}(2n) &=& \frac{1}{64\pi^4}\lim_{\alpha\to 0}\Bigg\{\nonumber\\
&& +\,\frac{1}{2}\sum_{\sigma_\alpha,\sigma_\beta}\int_0^\infty dq\,\frac{n_{\varepsilon_{q,\alpha}-i\sigma_\alpha \hat r\cdot\kappa}}{\varepsilon_{\alpha,q}}\int_0^\infty dk\,k\,\frac{n_{\varepsilon_{k,\beta}-i\sigma_\beta \hat r\cdot\lambda}}{\varepsilon_{\beta,k}}\,\left[\frac{\partial}{\partial q}\ln\big(\alpha\beta;\gamma\big)+\frac{4k}{\beta-\gamma}\right]\nonumber\\
&& +\,\frac{1}{2}\sum_{\sigma_\alpha,\sigma_\gamma}\int_0^\infty dq\,\frac{n_{\varepsilon_{q,\alpha}-i\sigma_\alpha \hat r\cdot\kappa}}{\varepsilon_{\alpha,q}}\int_0^\infty dk\,k\,\frac{n_{\varepsilon_{k,\gamma}-i\sigma_\gamma \hat r\cdot\tau}}{\varepsilon_{\gamma,k}}\,\left[\frac{\partial}{\partial q}\ln\big(\alpha\gamma;\beta\big)+\frac{4k}{\gamma-\beta}\right]\nonumber\\
&& -\,\sum_{\sigma_\beta,\sigma_\gamma}\int_0^\infty dq\,q\,\frac{n_{\varepsilon_{q,\beta}-i\sigma_\beta \hat r\cdot\lambda}}{\varepsilon_{\beta,q}}\int_0^\infty dk\,k\,\frac{n_{\varepsilon_{k,\gamma}-i\sigma_\gamma \hat r\cdot\tau}}{\varepsilon_{\gamma,k}}\,\frac{d}{d\alpha}\ln\big(\beta\gamma;\alpha\big)\Bigg\},
\eeq
where $\partial/\partial q$ denotes the partial derivative at $\varepsilon_{q,\alpha}$ fixed. Using
\beq
\frac{\partial}{\partial q}\ln\big(\alpha\beta;\gamma\big) & = & \frac{2(k+q)}{\gamma-\alpha-\beta-2\sigma_\alpha\sigma_\beta\varepsilon_{q,\alpha}\varepsilon_{k,\beta}+2qk}\nonumber\\
& & +\,\frac{2(k-q)}{\gamma-\alpha-\beta-2\sigma_\alpha\sigma_\beta\varepsilon_{q,\alpha}\varepsilon_{k,\beta}-2qk}\nonumber\\
& = & 4k\frac{\gamma-\alpha-\beta-2\sigma_\alpha\sigma_\beta\varepsilon_{q,\alpha}\varepsilon_{k,\beta}-2q^2}{(\gamma-\alpha-\beta-2\sigma_\alpha\sigma_\beta\varepsilon_{q,\alpha}\varepsilon_{k,\beta})^2-4q^2k^2}\,,
\eeq
as well as
\beq
\frac{d}{d\alpha}\ln\big(\beta\gamma;\alpha\big) & = & \frac{1}{\alpha-\beta-\gamma-2\sigma_\beta\sigma_\gamma\varepsilon_{q,\beta}\varepsilon_{k,\gamma}+2qk}\nonumber\\
& & -\,\frac{1}{\alpha-\beta-\gamma-2\sigma_\beta\sigma_\gamma\varepsilon_{q,\beta}\varepsilon_{k,\gamma}-2qk}\nonumber\\
& = & \frac{-4qk}{(\alpha-\beta-\gamma-2\sigma_\beta\sigma_\gamma\varepsilon_{q,\beta}\varepsilon_{k,\gamma})^2-4q^2k^2}\,,
\eeq
we find eventually
\beq
\Delta S_{0^2\beta\gamma}^{\kappa\lambda\tau}(2n) & = & \frac{1}{16\pi^4}\sum_{\sigma_\alpha,\sigma_\beta}\int_0^\infty dq\,n_{q-i\sigma_\alpha \hat r\cdot\kappa}\int_0^\infty dk\,k^2\,\frac{n_{\varepsilon_{k,\beta}-i\sigma_\beta \hat r\cdot\lambda}}{\varepsilon_{\beta,k}}\nonumber\\
& & \hspace{2.0cm}\times\,\frac{\sigma_\alpha\sigma_\beta \varepsilon_{k,\beta}+q\frac{\beta+\gamma}{\beta-\gamma}}{(\beta-\gamma)^2+4q^2\beta+4\sigma_\alpha\sigma_\beta (\beta-\gamma)q\varepsilon_{k,\beta}}\nonumber\\
& + & \frac{1}{16\pi^4}\sum_{\sigma_\alpha,\sigma_\gamma}\int_0^\infty dq\,n_{q-i\sigma_\alpha \hat r\cdot\kappa}\int_0^\infty dk\,k^2\,\frac{n_{\varepsilon_{k,\gamma}-i\sigma_\gamma \hat r\cdot\tau}}{\varepsilon_{\gamma,k}}\nonumber\\
& & \hspace{2.0cm}\times\,\frac{\sigma_\alpha\sigma_\gamma \varepsilon_{k,\gamma}+q\frac{\gamma+\beta}{\gamma-\beta}}{(\gamma-\beta)^2+4q^2\gamma+4\sigma_\alpha\sigma_\gamma (\gamma-\beta)q\varepsilon_{k,\gamma}}\nonumber\\
& + & \frac{1}{16\pi^4}\sum_{\sigma_\beta,\sigma_\gamma}\int_0^\infty dq\,q^2\int_0^\infty dk\,k^2\,\frac{n_{\varepsilon_{q,\beta}-i\sigma_\beta \hat r\cdot\lambda}n_{\varepsilon_{k,\gamma}-i\sigma_\gamma \hat r\cdot\tau}}{\varepsilon_{\beta,q}\varepsilon_{\gamma,k}}\nonumber\\
& & \hspace{1.0cm}\times\,\frac{1}{(\beta+\gamma)^2+4(\beta\gamma+q^2\gamma+k^2\beta)+4\sigma_\beta\sigma_\gamma(\beta+\gamma)\gamma\varepsilon_{q,\beta}\varepsilon_{k,\gamma}}\,.\label{eq:2n1}
\eeq
Note that the first two integrals remain safe when $\kappa=0$ provided we first sum over $\sigma_\alpha$.\\

\underline{$\Delta S^{\kappa\lambda\tau}_{0^20^2\gamma}(2n)$:} Similarly, the contribution with two thermal factors to $\Delta S^{\kappa\lambda\tau}_{0^20^2\gamma}$ reads
\beq
\Delta S_{0^20^2\gamma}^{\kappa\lambda\tau}(2n) &=& \frac{1}{64\pi^4}\lim_{\alpha\to 0}\lim_{\beta\to 0}\Bigg\{\nonumber\\
&& +\,\sum_{\sigma_\alpha,\sigma_\beta}\int_0^\infty dq\,q\,\frac{d}{d\alpha}\left(\frac{n_{\varepsilon_{q,\alpha}-i\sigma_\alpha \hat r\cdot\kappa}}{\varepsilon_{\alpha,q}}\right)\int_0^\infty dk\,k\,\frac{d}{d\beta}\left(\frac{n_{\varepsilon_{k,\beta}-i\sigma_\beta \hat r\cdot\lambda}}{\varepsilon_{\beta,k}}\right)\left[\ln\big(\alpha\beta;\gamma\big)-\frac{4qk}{\gamma}\right]\nonumber\\
& & +\,\sum_{\sigma_\alpha,\sigma_\beta}\int_0^\infty dq\,q\,\frac{d}{d\alpha}\left(\frac{n_{\varepsilon_{q,\alpha}-i\sigma_\alpha \hat r\cdot\kappa}}{\varepsilon_{\alpha,q}}\right)\int_0^\infty dk\,k\,\frac{n_{\varepsilon_{k,\beta}-i\sigma_\beta \hat r\cdot\lambda}}{\varepsilon_{\beta,k}}\left[\frac{d}{d\beta}\ln\big(\alpha\beta;\gamma\big)-\frac{4qk}{\gamma^2}\right]\nonumber\\
& & +\,\sum_{\sigma_\alpha,\sigma_\beta}\int_0^\infty dq\,q\,\frac{n_{\varepsilon_{q,\alpha}-i\sigma_\alpha \hat r\cdot\kappa}}{\varepsilon_{\alpha,q}}\int_0^\infty dk\,k\,\frac{d}{d\beta}\left(\frac{n_{\varepsilon_{k,\beta}-i\sigma_\beta \hat r\cdot\lambda}}{\varepsilon_{\beta,k}}\right)\left[\frac{d}{d\alpha}\ln\big(\alpha\beta;\gamma\big)-\frac{4qk}{\gamma^2}\right]\nonumber\\
& & +\,\sum_{\sigma_\alpha,\sigma_\gamma}\int_0^\infty dq\,q\,\frac{d}{d\alpha}\left(\frac{n_{\varepsilon_{q,\alpha}-i\sigma_\alpha \hat r\cdot\kappa}}{\varepsilon_{\alpha,q}}\right)\int_0^\infty dk\,k\,\frac{n_{\varepsilon_{k,\gamma}-i\sigma_\gamma \hat r\cdot\tau}}{\varepsilon_{\gamma,k}}\left[\frac{d}{d\beta}\ln\big(\alpha\gamma;\beta\big)+\frac{4qk}{\gamma^2}\right]\nonumber\\
& & +\,\sum_{\sigma_\beta,\sigma_\gamma}\int_0^\infty dq\,q\,\frac{n_{\varepsilon_{q,\gamma}-i\sigma_\gamma \hat r\cdot\tau}}{\varepsilon_{\gamma,q}}\int_0^\infty dk\,k\,\frac{d}{d\beta}\left(\frac{n_{\varepsilon_{k,\beta}-i\sigma_\beta \hat r\cdot\lambda}}{\varepsilon_{\beta,k}}\right)\left[\frac{d}{d\alpha}\ln\big(\gamma\beta;\alpha\big)+\frac{4qk}{\gamma^2}\right]\nonumber\\
& & +\,\sum_{\sigma_\alpha,\sigma_\beta}\int_0^\infty dq\,q\,\frac{n_{\varepsilon_{q,\alpha}-i\sigma_\alpha \hat r\cdot\kappa}}{\varepsilon_{\alpha,q}}\int_0^\infty dk\,k\,\frac{n_{\varepsilon_{k,\beta}-i\sigma_\beta \hat r\cdot\lambda}}{\varepsilon_{\beta,k}}\,\frac{d^2}{d\alpha d\beta}\ln\big(\alpha\beta;\gamma\big)\nonumber\\
& & +\,\sum_{\sigma_\alpha,\sigma_\gamma}\int_0^\infty dq\,q\,\frac{n_{\varepsilon_{q,\alpha}-i\sigma_\alpha \hat r\cdot\kappa}}{\varepsilon_{\alpha,q}}\int_0^\infty dk\,k\,\frac{n_{\varepsilon_{k,\gamma}-i\sigma_\gamma \hat r\cdot\tau}}{\varepsilon_{\gamma,k}}\,\frac{d^2}{d\alpha d\beta}\ln\big(\alpha\gamma;\beta\big)\nonumber\\
& & +\,\sum_{\sigma_\beta,\sigma_\gamma}\int_0^\infty dq\,q\,\frac{n_{\varepsilon_{q,\gamma}-i\sigma_\gamma \hat r\cdot\tau}}{\varepsilon_{\gamma,q}}\int_0^\infty dk\,k\,\frac{n_{\varepsilon_{k,\beta}-i\sigma_\beta \hat r\cdot\lambda}}{\varepsilon_{\beta,k}}\,\frac{d^2}{d\alpha d\beta}\ln\big(\gamma\beta;\alpha\big)\Bigg\}.
\eeq
Using integration by parts, we find
\beq
\Delta S_{0^20^2\gamma}^{\kappa\lambda\tau}(2n) &=& \frac{1}{64\pi^4}\lim_{\alpha\to 0}\lim_{\beta\to 0}\Bigg\{\nonumber\\
& & +\,\sum_{\sigma_\alpha,\sigma_\beta}\int_0^\infty dq\,q\,\frac{d}{d\alpha}\left(\frac{n_{\varepsilon_{q,\alpha}-i\sigma_\alpha \hat r\cdot\kappa}}{\varepsilon_{\alpha,q}}\right)\int_0^\infty dk\,k\,\frac{n_{\varepsilon_{k,\beta}-i\sigma_\beta \hat r\cdot\lambda}}{\varepsilon_{\beta,k}}\nonumber\\
& & \hspace{3.0cm}\times\,\left[\left(\frac{d}{d\beta}-\frac{1}{2}\frac{d}{dk^2}\right)\ln\big(\alpha\beta;\gamma\big)-\frac{4qk}{\gamma^2}+\frac{q}{k\gamma}\right]\nonumber\\
& & +\,\sum_{\sigma_\alpha,\sigma_\beta}\int_0^\infty dq\,q\,\frac{n_{\varepsilon_{q,\alpha}-i\sigma_\alpha \hat r\cdot\kappa}}{\varepsilon_{\alpha,q}}\int_0^\infty dk\,k\,\frac{d}{d\beta}\left(\frac{n_{\varepsilon_{k,\beta}-i\sigma_\beta \hat r\cdot\lambda}}{\varepsilon_{\beta,k}}\right)\nonumber\\
& & \hspace{3.0cm}\times\left[\left(\frac{d}{d\alpha}-\frac{1}{2}\frac{d}{dq^2}\right)\ln\big(\alpha\beta;\gamma\big)-\frac{4qk}{\gamma^2}+\frac{k}{q\gamma}\right]\nonumber\\
& & +\,\sum_{\sigma_\alpha,\sigma_\gamma}\int_0^\infty dq\,q\,\frac{d}{d\alpha}\left(\frac{n_{\varepsilon_{q,\alpha}-i\sigma_\alpha \hat r\cdot\kappa}}{\varepsilon_{\alpha,q}}\right)\int_0^\infty dk\,k\,\frac{n_{\varepsilon_{k,\gamma}-i\sigma_\gamma \hat r\cdot\tau}}{\varepsilon_{\gamma,k}}\left[\frac{d}{d\beta}\ln\big(\alpha\gamma;\beta\big)+\frac{4qk}{\gamma^2}\right]\nonumber\\
& & +\,\sum_{\sigma_\beta,\sigma_\gamma}\int_0^\infty dq\,q\,\frac{n_{\varepsilon_{q,\gamma}-i\sigma_\gamma \hat r\cdot\tau}}{\varepsilon_{\gamma,q}}\int_0^\infty dk\,k\,\frac{d}{d\beta}\left(\frac{n_{\varepsilon_{k,\beta}-i\sigma_\beta \hat r\cdot\lambda}}{\varepsilon_{\beta,k}}\right)\left[\frac{d}{d\alpha}\ln\big(\gamma\beta;\alpha\big)+\frac{4qk}{\gamma^2}\right]\nonumber\\
& & +\,\sum_{\sigma_\alpha,\sigma_\beta}\int_0^\infty dq\,q\,\frac{n_{\varepsilon_{q,\alpha}-i\sigma_\alpha \hat r\cdot\kappa}}{\varepsilon_{\alpha,q}}\int_0^\infty dk\,k\,\frac{n_{\varepsilon_{k,\beta}-i\sigma_\beta \hat r\cdot\lambda}}{\varepsilon_{\beta,k}}\,\frac{d^2}{d\alpha d\beta}\ln\big(\alpha\beta;\gamma\big)\nonumber\\
& & +\,\sum_{\sigma_\alpha,\sigma_\gamma}\int_0^\infty dq\,q\,\frac{n_{\varepsilon_{q,\alpha}-i\sigma_\alpha \hat r\cdot\kappa}}{\varepsilon_{\alpha,q}}\int_0^\infty dk\,k\,\frac{n_{\varepsilon_{k,\gamma}-i\sigma_\gamma \hat r\cdot\tau}}{\varepsilon_{\gamma,k}}\,\frac{d^2}{d\alpha d\beta}\ln\big(\alpha\gamma;\beta\big)\nonumber\\
& & +\,\sum_{\sigma_\beta,\sigma_\gamma}\int_0^\infty dq\,q\,\frac{n_{\varepsilon_{q,\gamma}-i\sigma_\gamma \hat r\cdot\tau}}{\varepsilon_{\gamma,q}}\int_0^\infty dk\,k\,\frac{n_{\varepsilon_{k,\beta}-i\sigma_\beta \hat r\cdot\lambda}}{\varepsilon_{\beta,k}}\,\frac{d^2}{d\alpha d\beta}\ln\big(\gamma\beta;\alpha\big)\Bigg\},
\eeq
and then
\beq
\Delta S_{0^20^2\gamma}^{\kappa\lambda\tau}(2n) &=& \frac{1}{64\pi^4}\lim_{\alpha\to 0}\lim_{\beta\to 0}\Bigg\{\nonumber\\
& & +\,\frac{1}{4}\sum_{\sigma_\alpha,\sigma_\beta}\int_0^\infty dq\,\frac{n_{\varepsilon_{q,\alpha}-i\sigma_\alpha \hat r\cdot\kappa}}{\varepsilon_{\alpha,q}}\int_0^\infty dk\,\frac{n_{\varepsilon_{k,\beta}-i\sigma_\beta \hat r\cdot\lambda}}{\varepsilon_{\beta,k}}\nonumber\\
& & \hspace{1.0cm}\times\,\left[\frac{\partial^2}{\partial q\partial k}\ln\big(\alpha\beta;\gamma\big)+\frac{8}{\gamma^2}(q^2+k^2)-\frac{4}{\gamma}\right]\nonumber\\
& & -\,\frac{1}{2}\sum_{\sigma_\alpha,\sigma_\gamma}\int_0^\infty dq\,\frac{n_{\varepsilon_{q,\alpha}-i\sigma_\alpha \hat r\cdot\kappa}}{\varepsilon_{\alpha,q}}\int_0^\infty dk\,k\,\frac{n_{\varepsilon_{k,\gamma}-i\sigma_\gamma \hat r\cdot\tau}}{\varepsilon_{\gamma,k}}\,\nonumber\\
& & \hspace{1.0cm}\times\,\left[\frac{\partial}{\partial q}\frac{d}{d\beta}\ln\big(\alpha\gamma;\beta\big)+\frac{4k}{\gamma^2}\right]\nonumber\\
& & -\,\frac{1}{2}\sum_{\sigma_\beta,\sigma_\gamma}\int_0^\infty dq\,q\,\frac{n_{\varepsilon_{q,\gamma}-i\sigma_\gamma \hat r\cdot\tau}}{\varepsilon_{\gamma,q}}\int_0^\infty dk\,\frac{n_{\varepsilon_{k,\beta}-i\sigma_\beta \hat r\cdot\lambda}}{\varepsilon_{\beta,k}}\,\nonumber\\
& & \hspace{1.0cm}\times\,\left[\frac{\partial}{\partial k}\frac{d}{d\alpha}\ln\big(\beta\gamma;\alpha\big)+\frac{4q}{\gamma^2}\right]\Bigg\}.
\eeq
We have
\beq
\frac{\partial^2}{\partial q\partial k}\ln\big(\alpha\beta;\gamma\big) & = & \frac{\partial}{\partial k}\frac{2(k+q)}{-(\sigma_\alpha\varepsilon_{q,\alpha}+\sigma_\beta\varepsilon_{k,\beta})^2+\varepsilon_{q+k,\gamma}^2}+\frac{\partial}{\partial k}\frac{2(k-q)}{-(\sigma_\alpha\varepsilon_{q,\alpha}+\sigma_\beta\varepsilon_{k,\beta})^2+\varepsilon_{q-k,\gamma}^2}\nonumber\\
& = &  \frac{2}{\gamma-\alpha-\beta-2\sigma_\alpha\sigma_\beta\varepsilon_{q,\alpha}\varepsilon_{k,\beta}+2qk}-\frac{4(k+q)^2}{(\gamma-\alpha-\beta-2\sigma_\alpha\sigma_\beta\varepsilon_{q,\alpha}\varepsilon_{k,\beta}+2qk)^2}\nonumber\\
& + & \frac{2}{\gamma-\alpha-\beta-2\sigma_\alpha\sigma_\beta\varepsilon_{q,\alpha}\varepsilon_{k,\beta}-2qk}-\frac{4(k-q)^2}{(\gamma-\alpha-\beta-2\sigma_\alpha\sigma_\beta\varepsilon_{q,\alpha}\varepsilon_{k,\beta}-2qk)^2}\,,\nonumber\\
\eeq
so, in the limit $\alpha\to 0$ and $\beta\to 0$,
\beq
& & \frac{\partial^2}{\partial q\partial k}\ln\big(\alpha\beta;\gamma\big)+\frac{8}{\gamma^2}(q^2+k^2)-\frac{4}{\gamma}\nonumber\\
& & \hspace{0.5cm}=\,\frac{2}{\gamma-X_-}-\frac{2}{\gamma}+\frac{2}{\gamma-X_+}-\frac{2}{\gamma}\nonumber\\
& & \hspace{0.5cm}+\,\frac{4}{\gamma^2}(q^2+k^2)-\frac{4(k+q)^2}{(\gamma-X_-)^2}+\frac{4}{\gamma^2}(q^2+k^2)-\frac{4(k-q)^2}{(\gamma-X_+)^2}\nonumber\\
& & \hspace{0.5cm}=\,\frac{2X_-}{\gamma(\gamma-X_-)}+\frac{4(q^2+k^2)(\gamma-X_-)^2-4(k+q)^2\gamma^2}{\gamma^2(\gamma-X_-)^2}\nonumber\\
& & \hspace{0.5cm}+\,\frac{2X_+}{\gamma(\gamma-X_+)}+\frac{4(q^2+k^2)(\gamma-X_+)^2-4(k-q)^2\gamma^2}{\gamma^2(\gamma-X_+)^2}\nonumber\\
& & \hspace{0.5cm}=\,\frac{2(2q^2+2k^2-\gamma)X_-^2-2(4q^2+4k^2-\gamma)\gamma X_--8qk\gamma^2}{\gamma^2(\gamma-X_-)^2}\nonumber\\
& & \hspace{0.5cm}+\,\frac{2(2q^2+2k^2-\gamma)X_+^2-2(4q^2+4k^2-\gamma)\gamma X_++8qk\gamma^2}{\gamma^2(\gamma-X_+)^2}\,,
\eeq
with $X_\pm\equiv 2(\sigma_\alpha\sigma_\beta\pm 1)qk$ such that $X_+X_-=0$, $X_++X_-=4\sigma_\alpha\sigma_\beta qk$, $X_+-X_-=4qk$ and $X_\pm^2=\pm 4qkX_\pm$. Using these properties, we find
\beq
& & \frac{\partial^2}{\partial q\partial k}\ln\big(\alpha\beta;\gamma\big)+\frac{8}{\gamma^2}(q^2+k^2)-\frac{4}{\gamma}\nonumber\\
& & \hspace{0.5cm}=\,\frac{-8qk(2q^2+2k^2-\gamma)X_--2(4q^2+4k^2-\gamma)\gamma X_--8qk\gamma^2}{\gamma^2(\gamma^2-(2\gamma+4qk) X_-)}\nonumber\\
& & \hspace{0.5cm}+\,\frac{8qk(2q^2+2k^2-\gamma)X_+-2(4q^2+4k^2-\gamma)\gamma X_++8qk\gamma^2}{\gamma^2(\gamma^2-(2\gamma-4qk)X_+)}\nonumber\\
& & \hspace{0.5cm}=\,\frac{(2\gamma^2-8\gamma(q^2+k^2-qk)-16qk(q^2+k^2))X_--8qk\gamma^2}{\gamma^2(\gamma^2-(2\gamma+4qk) X_-)}\nonumber\\
& & \hspace{0.5cm}+\,\frac{(2\gamma^2-8\gamma(q^2+k^2+qk)+16qk(q^2+k^2))X_++8qk\gamma^2}{\gamma^2(\gamma^2-(2\gamma-4qk)X_+)}\nonumber\\
& & \hspace{0.5cm}=\,\frac{\gamma^2(2\gamma^2-8\gamma(q^2+k^2-qk)-16qk(q^2+k^2))X_--8qk\gamma^2(\gamma^2-(2\gamma-4qk)X_+)}{\gamma^4(\gamma-4\sigma_\alpha\sigma_\beta qk)^2}\nonumber\\
& & \hspace{0.5cm}+\,\frac{\gamma^2(2\gamma^2-8\gamma(q^2+k^2+qk)+16qk(q^2+k^2))X_++8qk\gamma^2(\gamma^2-(2\gamma+4qk)X_-)}{\gamma^4(\gamma-4\sigma_\alpha\sigma_\beta qk)^2}\nonumber\\
& & \hspace{0.5cm}=\,8\sigma_\alpha\sigma_\beta \frac{qk}{\gamma^2}\frac{ \gamma^2-4\gamma(q^2+k^2)-16q^2k^2+4\sigma_\alpha\sigma_\beta qk(\gamma+2q^2+2k^2)}{\gamma^2+16q^2k^2-8\sigma_\alpha\sigma_\beta qk\gamma}\,.
\eeq
Similarly
\beq
\frac{d}{d\beta}\frac{\partial}{\partial q}\ln\big(\alpha\gamma;\beta) & = & \frac{d}{d\beta}\frac{2(k+q)}{-(\sigma_\alpha\varepsilon_{q,\alpha}+\sigma_\gamma\varepsilon_{k,\gamma})^2+\varepsilon_{q+k,\beta}^2}+\frac{d}{d\beta}\frac{2(k-q)}{-(\sigma_\alpha\varepsilon_{q,\alpha}+\sigma_\gamma\varepsilon_{k,\gamma})^2+\varepsilon_{q-k,\beta}^2}\nonumber\\
& = & -\frac{2(k+q)}{(\beta-\alpha-\gamma-2\sigma_\alpha\sigma_\gamma\varepsilon_{q,\alpha}\varepsilon_{k,\gamma}+2qk)^2}-\frac{2(k-q)}{(\beta-\alpha-\gamma-2\sigma_\alpha\sigma_\gamma\varepsilon_{q,\alpha}\varepsilon_{k,\gamma}-2qk)^2}\nonumber\\
& = & -4k\frac{(\beta-\alpha-\gamma-2\sigma_\alpha\sigma_\gamma\varepsilon_{q,\alpha}\varepsilon_{k,\gamma})^2+4q^2k^2-4q^2(\beta-\alpha-\gamma-2\sigma_\alpha\sigma_\gamma\varepsilon_{q,\alpha}\varepsilon_{k,\gamma})}{((\beta-\alpha-\gamma-2\sigma_\alpha\sigma_\gamma\varepsilon_{q,\alpha}\varepsilon_{k,\gamma})^2-4q^2k^2)^2}\,,\nonumber\\
\eeq
so, in the limit $\alpha\to 0$ and $\beta\to 0$,
\beq
& & \frac{d}{d\beta}\frac{\partial}{\partial q}\ln\big(\alpha\gamma;\beta)+\frac{4k}{\gamma^2}\nonumber\\
& & \hspace{0.5cm}=\,\frac{4k}{\gamma^2}-4k\frac{(\gamma+2\sigma_\alpha\sigma_\gamma q\varepsilon_{k,\gamma})^2+4q^2k^2+4q^2(\gamma+2\sigma_\alpha\sigma_\gamma q\varepsilon_{k,\gamma})}{((\gamma+2\sigma_\alpha\sigma_\gamma q\varepsilon_{k,\gamma})^2-4q^2k^2)^2}\nonumber\\
& & \hspace{0.5cm}=\,\frac{4k}{\gamma^2}-4k\frac{\gamma^2+8q^2\gamma+8q^2k^2+4\sigma_\alpha\sigma_\gamma q\varepsilon_{k,\gamma}(\gamma+2q^2)}{(\gamma^2+4q^2\gamma+4\sigma_\alpha\sigma_\gamma\gamma q\varepsilon_{k,\gamma})^2}\nonumber\\
& & \hspace{0.5cm}=\,\frac{4k}{\gamma^2}-4k\frac{\gamma^2+8q^2\gamma+8q^2k^2+4\sigma_\alpha\sigma_\gamma q\varepsilon_{k,\gamma}(\gamma+2q^2)}{\gamma^4+24q^2\gamma^3+16q^2(q^2+k^2)\gamma^2+8\sigma_\alpha\sigma_\gamma\gamma^2q\varepsilon_{k,\gamma}(\gamma+4q^2)}\nonumber\\
& & \hspace{0.5cm}=\,8qk\frac{8q\gamma+4q(2q^2+k^2)+2\sigma_\alpha\sigma_\gamma \varepsilon_{k,\gamma}(\gamma+6q^2)}{\gamma^4+24q^2\gamma^3+16q^2(q^2+k^2)\gamma^2+8\sigma_\alpha\sigma_\gamma\gamma^2q\varepsilon_{k,\gamma}(\gamma+4q^2)}\,.
\eeq
We deduce eventually that
\beq
\Delta S_{0^20^2\gamma}^{\kappa\lambda\tau}(2n) & = & \frac{2}{\gamma^2}\sum_{\sigma_\alpha,\sigma_\beta} \sigma_\alpha\sigma_\beta \int_0^\infty dq\,n_{q-i\sigma_\alpha \hat r\cdot\kappa}\int_0^\infty dk\,n_{k-i\sigma_\beta \hat r\cdot\lambda}\nonumber\\
& & \hspace{1.0cm}\times\,\frac{ \gamma^2-4\gamma(q^2+k^2)-16q^2k^2+4\sigma_\alpha\sigma_\beta qk(\gamma+2q^2+2k^2)}{\gamma^2+16q^2k^2-8\sigma_\alpha\sigma_\beta qk\gamma}\nonumber\\
& & -\,4\sum_{\sigma_\alpha,\sigma_\gamma}\int_0^\infty dq\,n_{q-i\sigma_\alpha \hat r\cdot\kappa}\int_0^\infty dk\,k^2\,\frac{n_{\varepsilon_{k,\gamma}-i\sigma_\gamma \hat r\cdot\tau}}{\varepsilon_{\gamma,k}}\,\nonumber\\
& & \hspace{1.0cm}\times\,\frac{8q\gamma+4q(2q^2+k^2)+2\sigma_\alpha\sigma_\gamma \varepsilon_{k,\gamma}(\gamma+6q^2)}{\gamma^4+24q^2\gamma^3+16q^2(q^2+k^2)\gamma^2+8\sigma_\alpha\sigma_\gamma\gamma^2q\varepsilon_{k,\gamma}(\gamma+4q^2)}\nonumber\\
& & -\,4\sum_{\sigma_\beta,\sigma_\gamma}\int_0^\infty dq\,q^2\,\frac{n_{\varepsilon_{q,\gamma}-i\sigma_\gamma \hat r\cdot\tau}}{\varepsilon_{\gamma,q}}\int_0^\infty dk\,n_{k-i\sigma_\beta \hat r\cdot\lambda}\,\nonumber\\
& & \hspace{1.0cm}\times\,\frac{8k\gamma+4k(2k^2+q^2)+2\sigma_\beta\sigma_\gamma \varepsilon_{q,\gamma}(\gamma+6k^2)}{\gamma^4+24k^2\gamma^3+16k^2(k^2+q^2)\gamma^2+8\sigma_\beta\sigma_\gamma\gamma^2k\varepsilon_{q,\gamma}(\gamma+4k^2)}\,.\label{eq:2n2}
\eeq

%%%
\subsection{Summary}
In summary, the subtracted simple and double mass derivatives of the scalar sunset sum-integral can be split as $\Delta S_{0^2\beta\gamma}^{\kappa\lambda\tau}=\Delta S_{0^2\beta\gamma}(0n)+\Delta S_{0^2\beta\gamma}^{\kappa\lambda\tau}(1n)+\Delta S_{0^2\beta\gamma}^{\kappa\lambda\tau}(2n)$ and $\Delta S_{0^20^2\gamma}^{\kappa\lambda\tau}=\Delta S_{0^20^2\gamma}(0n)+\Delta S_{0^20^2\gamma}^{\kappa\lambda\tau}(1n)+\Delta S_{0^20^2\gamma}^{\kappa\lambda\tau}(2n)$, with the vacuum contributions
\beq
\Delta S_{0^2\beta\gamma}(0n) & = & (d-3)\frac{\beta+\gamma}{(\beta-\gamma)^2}S_{0\beta\gamma}(0n)+(d-2)\frac{J_\beta(0n) J_\gamma(0n)}{(\beta-\gamma)^2}\,,\\
\Delta S_{0^20^2\gamma}(0n) & = & (d-3)(d-6)\frac{S_{00\gamma}(0n)}{\gamma^2}\,,
\eeq
the one thermal factor contributions
\beq
\Delta S^{\kappa\lambda\tau}_{0^2\beta\gamma}(1n) & = & J_{0}^\kappa(1n)\Delta I^{0^2}_{\beta\gamma}(0n)+J_\beta^\lambda(1n)\Delta I^\beta_{0^2\gamma}(0n)+J_\gamma^\tau(1n)\Delta I^\gamma_{0^2\beta}(0n)\,,\\
\Delta S^{\kappa\lambda\tau}_{0^20^2\gamma}(1n) & = &\big(J_{0}^\kappa(1n)+J_{0}^\lambda(1n)\big)\Delta I^{0^2}_{0^2\gamma}(0n)+J_{\gamma}^\tau(1n)\Delta I^\gamma_{0^20^2}(0n)\,,
\eeq
with
\beq
\Delta I^{0^2}_{\beta\gamma}(0n) & = & \frac{1}{(\beta-\gamma)^3}\left[\left(\gamma+\frac{4-d}{d}\beta\right)J_\beta(0n)-\left(\beta+\frac{4-d}{d}\gamma\right)J_\gamma(0n)\right],\\
\Delta I^\beta_{0^2\gamma}(0n) & = & \frac{1}{(\beta-\gamma)^2}\Big[(d-3)(\beta+\gamma)I^\beta_{0\gamma}(0n)+(d-2)J_\gamma(0n)\Big]\,,\\
\Delta I_{0^2\gamma}^{0^2}(0n) & = & 2\frac{d-6}{d}\frac{J_\gamma(0n)}{\gamma^3}\,,\\
\Delta I^\beta_{0^20^2}(0n) & = & (d-3)(d-6)\frac{I^\beta_{00}(0n)}{\beta^2}\,,
\eeq
and, finally, the two thermal factor contributions given in Eqs.~(\ref{eq:2n1}) and (\ref{eq:2n2}).

%%%%%
\section{Conclusions}
In this work, we have evaluated the scalar sunset diagrams with imaginary square masses that appear in the two-loop background potential in the GZ type model with a background gauge invariance from Ref. \cite{Kroff:2018ncl}. This also involves some mass derivatives of the scalar sunset in the limit where the corresponding mass is taken to zero. In fact, what appear are not mass derivatives by themselves, but specific combinations with tadpole integrals and their mass derivatives that admit regular limits in the zero mass limit. The evaluated cases include three scalar sunsets and three mass derivative combinations. In each case the square masses are either 0 or $\pm i m^2$. 

Through thermal splitting we have decomposed the sum-integrals into contributions with 0, 1 and 2 thermal factors. For the terms with 0 thermal factors, the vacuum contribution, we obtained the integral by analytic continuation from the results for real masses from Ref. \cite{Caffo:1998du}, for the cases where the non-vanishing masses were equal. In the other cases, i.e. the cases with square masses of opposite sign, we have made a direct evaluation adapting the technique of Ref. \cite{Caffo:1998du} to imaginary square masses.

 Instead of considering only the scalar sunset sum-integrals that appear in the GZ framework, one could make a broader study of all scalar sunset diagrams with purely imaginary masses. However, when using thermal splitting this requires some regularization of the denominator for the contributions with 1 or 2 thermal factors, in order to avoid singularities. This problem is particular for imaginary masses: in the case of real masses one simply adds a regulator to the denominator in the form of an infinitesimal imaginary number. In some cases, like the cases considered here, the imaginary masses themselves work as a regulator, making it impossible for the denominator to vanish, but this is not true in general. Even when we limit ourselves to cases where the square masses are either 0 or $\pm i m^2$, there are examples where the denominator can vanish, e.g. when one square mass is 0 and the other two square masses are $i m^2$. In principle, it is possible to find a consistent regularization for each case but one should investigate how this affects the subsequent steps of the calculation. Since this lies beyond the scope of the GZ application that we are pursuing, we leave this question for a future study.

The sunset diagrams that have been calculated in this work make up a substantial part of the calculation of the two-loop background potential in the GZ framework. We are currently evaluating the full two-loop potential \cite{vEgmond} in the presence of a temporal background in order to study the deconfinement transition in YM theory using this framework, as well as the interplay between the Polyakov loop and the Gribov parameter. It will be interesting to compare these results with the one-loop results in the same model from Ref.~\cite{Kroff:2018ncl}, as well as with the two-loop studies in the CF model at finite temperature from Refs.~\cite{Reinosa:2014ooa, Reinosa:2014zta,Reinosa:2015gxn,Reinosa:2015oua,Maelger:2017amh}.

\acknowledgements{D.~M. van Egmond wishes to acknowledge the hospitality from the {\it Centre de Physique Th\'eorique} at Ecole Polytechnique, where this work was developed.}

\appendix

%%%%%
\section{Regularization}\label{app:regularization}
After performing the spectral integrals in Eq. \eqref{eq:23}, one finds denominators of the form $\sigma_\alpha\varepsilon_{q,\alpha}+\sigma_\beta \varepsilon_{k,\beta}+\sigma_\gamma \varepsilon_{l,\gamma}$, with $\sigma_\alpha$, $\sigma_\beta$ and $\sigma_\gamma$ taking values in $\{-1,+1\}$, and $\smash{l=|\vec{q}+\vec{k}|}$. For real masses, one needs to add an imaginary regulator $i 0^+$ to the denominator to avoid divergences. For imaginary square masses, the discussion is more intricate. As we now argue, however, in all cases of interest for the GZ framework, a regulator is not necessary. To see when the denominators can vanish, we write
\beq
& & 0=\sigma_\alpha\varepsilon_{q,\alpha}+\sigma_\beta \varepsilon_{k,\beta}+\sigma_\gamma \varepsilon_{l,\gamma}\,,\nonumber\\
& \Leftrightarrow & (\sigma_\alpha\varepsilon_{q,\alpha}+\sigma_\beta \varepsilon_{k,\beta})^2=\varepsilon^2_{l,\gamma}\nonumber\\
& \Leftrightarrow & 2\sigma_\alpha\sigma_\beta \varepsilon_{q,\alpha}\varepsilon_{k,\beta}=\gamma-\alpha-\beta+l^2-q^2-k^2\nonumber\\
& \Leftrightarrow & 4\varepsilon^2_{q,\alpha}\varepsilon^2_{k,\beta}=(\gamma-\alpha-\beta+l^2-q^2-k^2)^2\nonumber\\
& \Leftrightarrow & \left\{\begin{array}{l}
0=R^2(\alpha,\beta,\gamma)+R^2(q^2,k^2,l^2)\\
0=q^2(\alpha-\beta-\gamma)+k^2(\beta-\gamma-\alpha)+l^2(\gamma-\alpha-\beta)
\end{array}\right.\,,
\eeq
where in the last step we have separated the condition into a real and an imaginary part, owing to the fact that $\alpha$, $\beta$ and $\gamma$ are purely imaginary. We recall that $l^2=q^2+k^2+2kq \cos\theta$, from which it follows that 
\beq
R^2(q^2,k^2,l^2) & = & q^4+k^4+(q^2+k^2)^2+4k^2q^2\cos^2\theta+4qk(q^2+k^2)\cos\theta\nonumber\\
& - & 2q^2k^2-2(q^2+k^2)(q^2+k^2+2qk\cos\theta)\nonumber\\
& = & -4q^2k^2\sin^2\theta\,.
\eeq
With this in mind, let us consider the cases of interest. We consider first the cases\footnote{The case $(\alpha,\beta,\gamma)=(im^2,-im^2,0)$ is relevant for the discussion of $\Delta S^{\kappa\lambda\tau}_{im^2(-im^2)0^2}$.} $(\alpha,\beta,\gamma)=(im^2,0,0)$, $(\alpha,\beta,\gamma)=(im^2,-im^2,0)$, and $(\alpha,\beta,\gamma)=(im^2,im^2,-im^2)$, for wich $R^2(\alpha,\beta,\gamma)$ equals $-m^4$, $-4m^4$ and $-5m^4$ respectively. In those cases, it is obvious that the first condition in (A1) cannot be satisfied unless $m=0$. Next, we consider $(\alpha,\beta,\gamma)=(im^2,im^2,im^2)$, for which $R^2(\alpha,\beta,\gamma)=3m^4$. In this case, the conditions (A1) read
\beq
0 & = & 3m^4-4q^2k^2\sin^2\theta\,,\\
0 & = & m^2(q^2+k^2+qk\cos\theta)\,.
\eeq
Since $m>0$, we can solve the second equation as $qk\cos\theta=-(q^2+k^2)$ and plug it back into the first condition to arrive at
\beq
0=3m^4+4(q^4+q^2k^2+k^4)\,,
\eeq
which has again no solution if $m>0$.

%%%%%
\section{Evaluation of $\tilde I_{\beta\gamma}(\varepsilon_{q,\alpha};{\bf q})$}\label{app:I}
Let us consider the vacuum Euclidean integral
\beq\label{eq:vac}
I_{\beta\gamma}(0n)(Q^2\geq 0)\equiv \int_K^{T=0}G_\beta(K)G_\gamma(L)\,,
\eeq
with imaginary square masses $\beta$ and $\gamma$. In order to make contact with $\tilde I_{\beta\gamma}(\varepsilon_{q,\alpha};{\bf q})$, let us evaluate the frequency integral in (\ref{eq:vac}) using the residue theorem. To this purpose, we write it as
\beq
I_{\beta\gamma}(0n)(Q^2\geq 0)\equiv\int_{{\bf k}} \int_{\cal C}\frac{dz}{2\pi i}\frac{1}{-z^2+\varepsilon_{k,\beta}^2}\frac{1}{-(z+iq_4)^2+\varepsilon_{l,\gamma}^2}\,,
\eeq
where the contour ${\cal C}$ is along the imaginary axis. Closing ${\cal C}$ on the right and noting that\footnote{Of course, the same result is obtained by closing the contour on the left.}
\beq
\frac{1}{-z^2+\varepsilon_{k,\beta}^2} & = & -\frac{1}{2\varepsilon_{k,\beta}}\left[\frac{1}{z-\varepsilon_{k,\beta}}-\frac{1}{z+\varepsilon_{k,\beta}}\right],\\
\frac{1}{-(z+iq_4)^2+\varepsilon_{l,\gamma}^2} & = & -\frac{1}{2\varepsilon_{l,\gamma}}\left[\frac{1}{z+iq_4-\varepsilon_{l,\gamma}}-\frac{1}{z+iq_4+\varepsilon_{l,\gamma}}\right],
\eeq
we find
\beq\label{eq:eucl_res}
I_{\beta\gamma}(0n)(Q^2\geq 0) & = & \int_{{\bf k}} \left[\frac{1}{2\varepsilon_{k,\beta}}\frac{1}{-(\varepsilon_{k,\beta}+iq_4)^2+\varepsilon_{l,\gamma}^2}+\frac{1}{2\varepsilon_{l,\gamma}}\frac{1}{-(\varepsilon_{l,\gamma}-iq_4)^2+\varepsilon_{k,\beta}^2}\right]\nonumber\\
& = & \int_{{\bf k}} \frac{1}{4\varepsilon_{k,\beta}\varepsilon_{l,\gamma}}\left[\frac{1}{\varepsilon_{k,\beta}+\varepsilon_{l,\gamma}+iq_4}+\frac{1}{\varepsilon_{k,\beta}+\varepsilon_{l,\gamma}-iq_4}\right]\nonumber\\
& = & \int_{{\bf k}} \frac{1}{2\varepsilon_{k,\beta}\varepsilon_{l,\gamma}}\frac{\varepsilon_{k,\beta}+\varepsilon_{l,\gamma}}{(\varepsilon_{k,\beta}+\varepsilon_{l,\gamma})^2-\varepsilon_{q,-Q^2}^2}\,.
\eeq
In the last step, we have written $q_4^2$ as $-\varepsilon^2_{q,-Q^2}$ to emphasize the fact this last integral is similar to the one defining $\tilde I_{\beta\gamma}(\varepsilon_{q,\alpha};{\bf q})$ in Eq.~(\ref{eq:tildeI}), with $\alpha$ replaced by $-Q^2$. More precisely, if we introduce the function 
\beq
F_{\bf q}(Q^2)\equiv\int_{{\bf k}} \frac{1}{2\varepsilon_{k,\beta}\varepsilon_{l,\gamma}}\frac{\varepsilon_{k,\beta}+\varepsilon_{l,\gamma}}{(\varepsilon_{k,\beta}+\varepsilon_{l,\gamma})^2-\varepsilon_{q,-Q^2}}\,,
\eeq
seen now as a function of a complex $Q^2$ for a fixed ${\bf q}$, we have both $F_{\bf q}(Q^2\geq 0)=I_{\beta\gamma}(0n)(Q^2\geq 0)$ and $F_{\bf q}(Q^2=-\alpha\in i\mathds{R})=\tilde I_{\beta\gamma}(\varepsilon_{q,\alpha};{\bf q})$.

%%%
\subsection{Analytic continuation}
To turn this observation into a practical way to determine $\tilde I_{\beta\gamma}(\varepsilon_{q,\alpha};{\bf q})$, we note first that if $\tilde I_{\beta\gamma}(\varepsilon_{q,\alpha};{\bf q})$ makes sense for a given value of $\alpha$, it makes sense for any other value close to it. In particular, we can write 
\beq
\tilde I_{\beta\gamma}(\varepsilon_{q,\alpha};{\bf q})=F_{\bf q}(Q^2=-\alpha+0^+)\,.
\eeq
Second, it is easily seen that $F_{\bf q}(Q^2)$ is analytic in the semi-plane ${\rm Re}\,Q^2>0$. Indeed, the potential singularities are restricted to the region defined by the condition $(\varepsilon_{k,\beta}+\varepsilon_{l,\gamma})^2-\varepsilon^2_{q,-Q^2}=0$, which corresponds to
\beq
Q^2=-(k^2+l^2-q^2+2\varepsilon_{k,\beta}\varepsilon_{l,\gamma})-\beta-\gamma\,,
\eeq
and whose real part obeys
\beq
{\rm Re}\,Q^2\leq-(k^2+l^2-q^2+2|k||l|)\leq -((|k|+|l|)^2-(k+l)^2)\leq 0\,.
\eeq
From these considerations, it follows that, if we know explicitly a function $G(Q^2)$ which is analytic over an open connected subset $\Omega$ of ${\rm Re}\,Q^2>0$ containing both the $Q^2=-\alpha+0^+$ and the $Q^2>0$ axis, and which agrees with $I_{\beta\gamma}(0n)(Q^2\geq 0)$ along this axis, then $G(Q^2)=F_q(Q^2)$ over $\Omega$. In particular, $\tilde I_{\beta\gamma}(\varepsilon_{q,\alpha};{\bf q})$ can be obtained as $G(Q^2=-\alpha+0^+)$.

%%%
\subsection{Explicit expression}
It remains to construct explicit examples of $G(Q^2)$. This is easily done by evaluating $I_{\beta\gamma}(0n)(Q^2\geq 0)$ using the Feynman trick and by extending the result to complex values of $Q^2$. Of course, as long as $Q^2\geq 0$, there are many equivalent forms of $G(Q^2)$ that one can write. In order to determine $\tilde I_{\beta\gamma}(\varepsilon_{q,\alpha};{\bf q})$, we should only use those forms that obey the above mentioned analytic properties.

The Feynman trick allows us to write
\beq
I_{\beta\gamma}(0n)(Q^2\geq 0)=\frac{1}{16\pi^2}\left\{\frac{1}{\epsilon}+\ln\frac{\bar\mu^2}{Q^2}-\int_0^1dx \ln\left(x(1-x)+x\frac{\beta}{Q^2}+(1-x)\frac{\gamma}{Q^2}\right)\right\}.\nonumber\\
\eeq
To perform the integral over $x$, it is convenient to write
\beq
& & x(1-x)+x\frac{\beta}{Q^2}+(1-x)\frac{\gamma}{Q^2}\nonumber\\
& & \hspace{0.5cm} =\,-x^2+\left(1+\frac{\beta-\gamma}{Q^2}\right)x+\frac{\gamma}{Q^2}\nonumber\\
& & \hspace{0.5cm} =\,-\left(x-\frac{1}{2}\left(1+\frac{\beta-\gamma}{Q^2}\right)\right)^2+\frac{\gamma}{Q^2}+\frac{1}{4}\left(1+\frac{\beta-\gamma}{Q^2}\right)^2\nonumber\\
& & \hspace{0.5cm}=\,\left(\frac{R(-Q^2,\beta,\gamma)+Q^2+\beta-\gamma}{2Q^2}-x\right)\left(\frac{R(-Q^2,\beta,\gamma)-Q^2-\beta+\gamma}{2Q^2}+x\right),\label{eq:factors}
\eeq
with 
\beq
R^2(-Q^2,\beta,\gamma) & = & Q^4+\beta^2+\gamma^2+2Q^2(\beta+\gamma)-2\beta\gamma\nonumber\\
& = & Q^4+2Q^2(\beta+\gamma)+(\beta-\gamma)^2\nonumber\\
& = & (Q^2+\beta+\gamma)^2-4\beta\gamma\,.\label{eq:R2}
\eeq
If we want to split the logarithm of the product of the two factors in Eq.~(\ref{eq:factors}), we need to check that the sum of the arguments of the factors lies between $-\pi$ and $\pi$. Since, the product of the factors never crosses the branch cut, it is enough to work at $x=0$ and $Q^2=0$. One finds
\beq
& & {\rm Arg}\big(R(0,\beta,\gamma)+\beta-\gamma\big)+{\rm Arg}\big(R(0,\beta,\gamma)-\beta+\gamma\big)\nonumber\\
& & \hspace{0.7cm}=\,{\rm Arg}\big(\sqrt{(\beta-\gamma)^2}+(\beta-\gamma)\big)+{\rm Arg}\big(\sqrt{(\beta-\gamma)^2}-(\beta-\gamma)\big)\nonumber\\
& & \hspace{0.7cm}=\,{\rm Arg}(2i{\rm Max}(-i\beta,-i\gamma))=\frac{\pi}{2}{\rm sign}({\rm Max}(-i\beta,-i\gamma))\,,\label{eq:discussion}
\eeq
which lies between $-\pi/2$ and $\pi/2$. We can then split the logarithm and compute the $x$-integral to obtain, after some trivial simplifications,
\beq
& & I_{\beta\gamma}(0n)(Q^2\geq 0)=G_{(0)}(Q^2)\nonumber\\
& & \hspace{1.0cm}\equiv\,\frac{1}{16\pi^2}\left\{\frac{1}{\epsilon}+\ln\frac{\bar\mu^2}{Q^2}+2\right.\nonumber\\
& & \hspace{3.2cm}-\,\frac{R(-Q^2,\beta,\gamma)+Q^2-\beta+\gamma}{2Q^2}\ln\frac{R(-Q^2,\beta,\gamma)+Q^2-\beta+\gamma}{2Q^2}\nonumber\\
& & \hspace{3.2cm}+\,\frac{R(-Q^2,\beta,\gamma)-Q^2-\beta+\gamma}{2Q^2}\ln\frac{R(-Q^2,\beta,\gamma)-Q^2-\beta+\gamma}{2Q^2}\nonumber\\
& & \hspace{3.2cm}+\,\frac{R(-Q^2,\beta,\gamma)-Q^2+\beta-\gamma}{2Q^2}\ln\frac{R(-Q^2,\beta,\gamma)-Q^2+\beta-\gamma}{2Q^2}\nonumber\\
& & \hspace{3.2cm}\left.-\,\frac{R(-Q^2,\beta,\gamma)+Q^2+\beta-\gamma}{2K^2}\ln\frac{R(-Q^2,\beta,\gamma)+Q^2+\beta-\gamma}{2Q^2}\right\}\!.
\eeq
It will be convenient to rewrite this Euclidean expression as
\beq\label{eq:eucl}
& & I_{\beta\gamma}(0n)(Q^2\geq 0)=G_{(1)}(Q^2)\nonumber\\
& & \hspace{1.0cm}\equiv\,\frac{1}{16\pi^2}\left\{\frac{1}{\epsilon}-\ln\frac{\bar\mu^2}{Q^2}+2\right.\nonumber\\
& & \hspace{3.2cm}-\,\frac{R(-Q^2,\beta,\gamma)+Q^2-\beta+\gamma}{2Q^2}\ln\frac{R(-Q^2,\beta,\gamma)+Q^2-\beta+\gamma}{2\bar\mu^2}\nonumber\\
& & \hspace{3.2cm}+\,\frac{R(-Q^2,\beta,\gamma)-Q^2-\beta+\gamma}{2Q^2}\ln\frac{R(-Q^2,\beta,\gamma)-Q^2-\beta+\gamma}{2\bar\mu^2}\nonumber\\
& & \hspace{3.2cm}+\,\frac{R(-Q^2,\beta,\gamma)-Q^2+\beta-\gamma}{2Q^2}\ln\frac{R(-Q^2,\beta,\gamma)-Q^2+\beta-\gamma}{2\bar\mu^2}\nonumber\\
& & \hspace{3.2cm}\left.-\,\frac{R(-Q^2,\beta,\gamma)+Q^2+\beta-\gamma}{2Q^2}\ln\frac{R(-Q^2,\beta,\gamma)+Q^2+\beta-\gamma}{2\bar\mu^2}\right\}\!,
\eeq
where we have changed the scale under the logarithms to the price of changing $\ln\bar\mu^2/Q^2$ by $-\ln\bar\mu^2/Q^2$.

Let us now analyse the singularities of $G_{(1)}(Q^2)$. We mention that we are not after the precise determination of the singularities. Rather we want to check that they comply with the above mentioned requirements, allowing one to extract $I_{\beta\gamma}(\varepsilon_{q,\alpha};{\bf q})$ as $G_{(1)}(Q^2=-\alpha+0^+)$. First, there is a  branch cut along the negative real axis, originating from $\ln\bar\mu^2/Q^2$. A second branch cut originates from $R(-Q^2,\beta,\gamma)$ which contains a square root. More precisely, this branch cut corresponds to
\beq
R^2(-Q^2_{\sqrt{\phantom{}}},\beta,\gamma)=-u,\,\,\mbox{with } u>0\,.
\eeq
This is easily solved using the third form of $R^2(-Q^2,\beta,\gamma)$ in (\ref{eq:R2}), and we find
\beq
Q^2_{\sqrt{\phantom{}}}=-\beta-\gamma\pm\sqrt{4\beta\gamma-u}\,\,\mbox{with } u>0\,.
\eeq
Finally, from the logarithms, we have potentially four branch cuts corresponding to
\beq
Q^2_{\ln1}+(\beta-\gamma)\pm R(-Q^2_{\ln1},\beta,\gamma) & = & -u\,\, \mbox{with } u>0\,,\\
Q^2_{\ln2}-(\beta-\gamma)\pm R(-Q^2_{\ln2},\beta,\gamma) & = & -u\,\, \mbox{with } u>0\,.
\eeq
Using the first form of $R^2(-Q^2,\beta,\gamma)$ in (\ref{eq:R2}), we find
\beq
Q^2_{\ln1} & = & -\frac{u}{2}\frac{u+2(\beta-\gamma)}{u-2\gamma}=-\frac{u}{2}\frac{u^2+2\beta u+4(\beta-\gamma)\gamma}{u^2-4\gamma^2}\,,\\
Q^2_{\ln2} & = & -\frac{u}{2}\frac{u-2(\beta-\gamma)}{u-2\beta}=-\frac{u}{2}\frac{u^2+2\gamma u+4(\gamma-\beta)\beta}{u^2-4\beta^2}\,.
\eeq
It is easily checked that, even though there are some singularities in the semi-plane ${\rm Re}\,Q^2>0$, they comply with the requirements and therefore
\beq
\tilde I_{\beta\gamma}(\varepsilon_{q,\alpha};{\bf q})=G_{(1)}(Q^2\to -\alpha+0^+)\,.
\eeq
We note in particular that $\tilde I_{\beta\gamma}(\varepsilon_{q,\alpha};{\bf q})$ does not depend on ${\bf q}$. This could have been anticipated from the fact that the analytic continuation is unique.

%%%%%
\section{Evaluation of $S_{\alpha\alpha(-\alpha)}(0n)$}\label{app:S++-}
Let us consider the slightly more general quantity $S_{\alpha\alpha\gamma}(0n)$. Without loss of generality, we can assume that $\alpha=im^2$ with $m^2>0$. However, we take $\gamma=ic^2$, with $c^2\in\mathds{R}$. Writing $S_{\alpha\alpha\gamma}(0n)\equiv\bar V(m^2, c^2)/(2\pi)^4$ and generalizing the argumentation of \cite{Caffo:1998du}, we write
\beq\label{eq:laporta}
\bar V(m^2,c^2)=(4\pi\mu^2)^{4-d}\Gamma\left(3-\frac{d}{2}\right)^2\left[\frac{\bar V^{(-2)}(m^2,c^2)}{(d-4)^2}+\frac{\bar V^{(-1)}(m^2,c^2)}{d-4}+\bar V^{(0)}(m^2,c^2)+\dots\right]\!.\nonumber\\
\eeq
Each of the $\bar V^{(j)}$'s obeys a differential equation that can be derived from Eq.~(\ref{eq:nice}). One finds
\beq
R^2(m^2,c^2)\frac{\partial}{\partial c^2} \bar V^{(j)}(m^2, c^2)=(c^2-2m^2)\bar V^{(j)}(m^2, c^2)+\bar{g}^{(j)}(m^2,c^2)\,,\label{eq:diff}
\eeq
with $R^2(m^2,c^2)\equiv R^2(m^2,m^2,c^2)=c^2(c^2-4m^2)$ and
\beq
\bar g^{(-2)}(m^2,c^2) & = & \frac{i}{2}m^2(c^2-m^2)\,,\label{eq:gm2}\\
\bar g^{(-1)}(m^2,c^2) & = & (c^2-2m^2)\bar V^{(-2)}(m^2,c^2)\nonumber\\
& + & \frac{im^2}{4}\Big[m^2-c^2+(c^2-2m^2)\ln (im^2)+c^2\ln(ic^2)\Big]\,,\label{eq:gm1}\\
\bar g^{(0)}(m^2,c^2) & = & (c^2-2m^2)\bar V^{(-1)}(m^2,c^2)\nonumber\\
& + & \frac{im^2}{8}\Bigg[c^2-m^2+\left(\frac{c^2}{2}-2m^2\right)\ln^2(im^2)+\frac{c^2}{2}\ln^2(ic^2)+\nonumber\\
&  & \hspace{0.5cm}+\,c^2\ln(ic^2)\ln(im^2)+(2m^2-c^2)\ln(im^2)-c^2\ln(ic^2)\Bigg].\label{eq:gm0}
\eeq 
The determination of the $\bar V^{(j)}$'s proceeds recursively: one first determines $\bar V^{(-2)}$ by integrating the corresponding differential equation with the explicit expression (\ref{eq:gm2}) for $\bar g^{(-2)}$. Knowing $\bar V^{(-2)}$, one can then determine $\bar g^{(-1)}$ from (\ref{eq:gm1}) and repeat the procedure, until all the $\bar V^{(j)}$'s have been determined. We mention that the integration of each differential equation gives each $\bar V^{(j)}(m^2,c^2)$, in terms of a boundary value $\bar V^{(j)}(m^2,c_0^2)$. There seems to be a circular reasoning {\it a priori}. We see below how this problem is avoided.

%%%
\subsection{Integrating the differential equation and boundary value}
Each differential equation (\ref{eq:diff}) is valid separately over $c^2<0$, $0<c^2<4m^2$ and $c^2>4m^2$. We here focus on the regions $c^2<0$ and $c^2>4m^2$, in which case $R^2(m^2,c^2)>0$. Following \cite{Caffo:1998du}, we note that
\beq
\frac{\partial}{\partial c^2}\frac{1}{R(m^2,c^2)} & = & -\frac{c^2-2m^2}{R^3(m^2,c^2)}\,,\\
\frac{\partial}{\partial c^2}\frac{c^2-2m^2}{R(m^2,c^2)} & = & -\frac{4m^4}{R^3(m^2,c^2)}\,.
\eeq
It follows that the differential equation can be rewritten as
\beq
\frac{\partial}{\partial c^2}\left(\frac{4m^4\bar V^{(j)}(m^2,c^2)+\bar{g}^{(j)}(c^2,m^2)(c^2-2m^2)}{R(m^2,c^2)}\right)=\frac{c^2-2m^2}{R(m^2,c^2)}\frac{\partial\bar{g}^{(j)}(c^2,m^2)}{\partial c^2}\,.
\eeq
The benefit of this rewriting is two-fold. First, it can be integrated to provide an expression for $\bar V^{(j)}(m^2,c^2)$ in terms of an integral involving $\bar g^{(j)}(m^2,c^2)$ and a boundary value $\bar V^{(j)}(m^2,c_0^2)$. Second, by choosing $c_0^2=0$ or $c_0^2=4m^4$ (depending on the considered region), this boundary is not needed because
\beq
4m^4\bar V^{(j)}(m^2,c^2_0)+\bar{g}^{(j)}(c^2_0,m^2)(c^2_0-2m^2)=0\,,
\eeq
owing to Eq.~(\ref{eq:diff}). It follows that
\beq\label{eq:sol}
\bar V^{(j)}(m^2,c^2) & = & \frac{1}{4m^4}\left[R(m^2,c^2)\int_{c^2_0}^{c^2}dx\,\frac{x-2m^2}{R(m^2,x)}\frac{\partial\bar{g}^{(j)}(m^2,x)}{\partial x}+(2m^2-c^2)\bar{g}^{(j)}(m^2,c^2)\right],\nonumber\\
\eeq
which provides a one-dimensional integral representation for $\bar V^{(j)}(m^2,c^2)$.

%%%
\subsection{Computing the remaining integrals}
It remains to evaluate the integral in Eq.~(\ref{eq:sol}). To this purpose, it is convenient to consider the change of variables
\beq
x=m^2\frac{(t+1)^2}{t}\,,\label{eq:cov}
\eeq
such that
\beq
\frac{dx}{dt}=m^2\left(1-\frac{1}{t^2}\right).
\eeq
The function $x(t)$ increases from $x(-\infty)=-\infty$ to $x(-1)=0$ and then decreases to $x(0^-)=-\infty$. Similarly, it decreases from $x(0^+)=+\infty$ to $x(1)=4m^2$ and then increases to $x(+\infty)=+\infty$. This means that the change of variables (\ref{eq:cov}) is adapted to the regions $x<0$ and $x>4m^2$. In each case, there are two possible branches obtained by solving a quadratic equation whose discriminant is $\Delta=(2m^2-x^2)^2-4m^4=R^2(m^2,x^2)$. The branches read
\beq
t=t_\pm(x)\equiv\frac{x-2m^2\pm R(m^2,x)}{2m^2}\,,
\eeq
with $0<t_-(x)<1<t_+(x)$ in the case where $x>4 m^2$, whereas $t_-(x)<-1<t_+(x)<0$ in the case where $x<0$. We note for later purpose that 
\beq
t_+t_-=1, \quad  t_++t_-=\frac{x^2}{m^2}-\frac{1}{2}, \quad t_+-t_-=\frac{R(m^2,x^2)}{m^2}\,.
\eeq 
Moreover, if we choose to work with $t_\tau(x)$, with $\tau=\pm 1$, then
\beq
R(m^2,x)=\tau m^2\left(2t-\frac{t^2+1}{t}\right)=\tau m^2\frac{t^2-1}{t}\,,
\eeq
It follows that Eq.~(\ref{eq:sol}) rewrites
\beq
\bar V^{(j)}(m^2,c^2) & = & \frac{1}{4m^2}\left[\tau  R(m^2,c^2)\int_{{\rm sgn}(c^2)}^{t_\tau(c^2)}dt\,\frac{t^2+1}{t^2}\frac{\partial\bar{g}^{(j)}(x)}{\partial x}+(2m^2-c^2)\frac{\bar{g}^{(j)}(m^2,c^2)}{m^2}\right]\!,\label{eq:int_final}
\eeq
and the result should not depend on the value of $\tau$. 

It is now easy to see that the procedure described below Eqs.~(\ref{eq:gm2})-(\ref{eq:gm0}) generates the following integrals
\beq
I_0 & \equiv & \int_{\sigma}^{t_\tau} dt\,\frac{1+t^2}{t^2}\,,\\
I_1 & \equiv & \int_{\sigma}^{t_\tau} dt\,\frac{1+t^2}{t^2}\,x\,,\\
J_0 & \equiv & \int_{\sigma}^{t_\tau} dt\,\frac{1+t^2}{t^2}\,\ln(ix)\,,\\
J_1 & \equiv & \int_{\sigma}^{t_\tau} dt\,\frac{1+t^2}{t^2}\,x\,\ln(ix)\,,\\
K_0 & \equiv & \frac{1}{2}\int_{\sigma}^{t_\tau} dt\,\frac{1+t^2}{t^2}\,\ln^2(ix)\,,
\eeq
where we have introduced $\sigma\equiv{\rm sgn}(c^2)$ and $t_\tau\equiv t_\tau(c^2)$ for simplicity. For the first one, we have
\beq
I_0=\left[t-\frac{1}{t}\right]_{\sigma}^{t_\tau}=t_\tau-t_{-\tau}=\tau\frac{R(m^2,c^2)}{m^2}\,.
\eeq
Similarly
\beq
I_1 & = & 2m^2\int_{\sigma}^{t_\tau} dt\,\frac{1+t^2}{t^2}+m^2\int_{\sigma}^{t_\tau} dt\,\frac{(1+t^2)^2}{t^3}\nonumber\\
& = & 2m^2\int_{\sigma}^{t_\tau} dt\,\frac{1+t^2}{t^2}+m^2\int_{\sigma}^{t_\tau} dt\,\left(t+\frac{1}{t^3}+\frac{2}{t}\right)\nonumber\\
& = & 2m^2I_0+m^2\left[\frac{1}{2}\left(t^2_\tau-\frac{1}{t^2_\tau}\right)+2\ln|t_\tau|\right]\nonumber\\
& = & 2\tau R(m^2,c^2)+\left[\tau R(m^2,c^2)\left(\frac{c^2}{2m^2}-1\right)+2m^2\ln|t_\tau|\right]\nonumber\\
& = &  \tau R(m^2,c^2)\left(1+\frac{c^2}{2m^2}\right)+2m^2\ln|t_\tau|\,.
\eeq
To treat the other integrals, we use the formula
\beq
\int_\sigma^{t_\tau}dt\,f'(t)\,\ln(ix) & = & f(t)\ln(ix)\Big|_\sigma^{t_\tau} -\,\int_\sigma^{t_\tau}dt\,f(t)\frac{t-1}{t (t+1)}\,,
\eeq
obtained via integration by parts. In particular, we have
\beq
J_0 & = & \tau\frac{R(m^2,c^2)}{m^2}\ln (ic^2)-\int_{\sigma}^{t_\tau}dt\,\frac{(t-1)^2}{t^2}\nonumber\\
& = & \tau\frac{R(m^2,c^2)}{m^2}\ln (ic^2)-\int_{\sigma}^{t_\tau}dt\,\left(1+\frac{1}{t^2}-\frac{2}{t}\right)\nonumber\\
& = &  \tau\frac{R(m^2,c^2)}{m^2}\Big(\ln (ic^2)-1\Big)+2\ln|t_\tau|\,,
\eeq
and, similarly,
\beq
J_1 & = & 2m^2\int_{\sigma}^{t_\tau} dt\,\frac{1+t^2}{t^2}\ln (ix)+m^2\int_{\sigma}^{t_\tau} dt\,\frac{(1+t^2)^2}{t^3}\ln (ix)\nonumber\\
& = & 2m^2J_0+m^2\int_{\sigma}^{t_\tau} dt\,\left(t+\frac{1}{t^3}\right)\ln (ix)+2m^2\int_{\sigma}^{t_\tau} dt\,\frac{\ln (ix)}{t}\nonumber\\
& = & 2m^2J_0+\tau R(m^2,c^2)\left(\frac{c^2}{2m^2}-1\right)\ln (ic^2)\nonumber\\
& - & \frac{m^2}{2}\int_{\sigma}^{t_\tau} dt\,\left(t+\frac{1}{t^3}-2-\frac{2}{t^2}+\frac{2}{t}\right)+2m^2\int_{\sigma}^{t_\tau} dt\,\frac{\ln (ix)}{t}\nonumber\\
& = & 2m^2J_0+\tau R(m^2,c^2)\left(\frac{c^2}{2m^2}-1\right)\ln (ic^2)+\frac{1}{2}\tau R(m^2,c^2)\left(3-\frac{c^2}{2m^2}\right)\\
& - & m^2 \ln|t_\tau|+2m^2\int_{\sigma}^{t_\tau} dt\,\frac{\ln (ix)}{t}\nonumber\\
& = & \tau R(m^2,c^2)\left(\frac{c^2}{2m^2}+1\right)\ln (ic^2)-\frac{1}{2}\tau R(m^2,c^2)\left(1+\frac{c^2}{2m^2}\right)\nonumber\\
& + & 3m^2 \ln|t_\tau|+2m^2\int_{\sigma}^{t_\tau} dt\,\frac{\ln (ix)}{t}\,.
\eeq
Using similar ideas, we find
\beq
K_0 & = & \tau \frac{R(m^2,c^2)}{m^2}\frac{\ln^2 (ic^2)}{2}-\int_{\sigma}^{t_\tau} dt\,\left(1+\frac{1}{t^2}-\frac{2}{t}\right)\ln (ix)\nonumber\\
& = & \tau\frac{R(m^2,c^2)}{m^2}\frac{\ln^2 (ic^2)}{2}-J_0+2\int_{\sigma}^{t_\tau}dt\,\frac{\ln (ix)}{t}\,.
\eeq
In these last two expressions, we need the integral
\beq
\int_{\sigma}^{t_\tau}dt\,\frac{\ln (ix)}{t} & =& \int_{\sigma}^{t_\tau}dt\,\frac{\ln m^2}{t}+2 \int_{-\sigma}^{-t_\tau}dt\,\frac{\ln |1-t|}{t}-\int_{\sigma}^{t_\tau}dt\,\frac{\ln |t|}{t}+\int_{\sigma}^{t_\tau}dt\,\frac{ \ln (\rm{sgn(t)} i)}{t}\nonumber\\
& = & \ln m^2 \ln |t_\tau|-2 \Phi (-t_\tau)+2\, \Phi(-\sigma)-\frac{1}{2}\ln^2|t_\tau| + i \frac{\pi}{2}\sigma\ln |t_\tau|\,,
\eeq
where we have introduced Spence function
\beq
\Phi(u)=-\int_0^u\frac{dx}{x}\ln|1-x|\,.
\eeq
We note that
\beq
\Phi({\rm sgn}(u)) & = & \sum_{k=1}^\infty \int_0^{{\rm sgn}(u)} dx\,\frac{x^{k-1}}{k}=\sum_{k=1}^\infty \frac{({\rm sgn}(u))^k}{k^2}\nonumber\\
& = & \sum_{k\,{\rm even}}\frac{1}{k^2}+{\rm sgn(u)}\sum_{k\,{\rm odd}}\frac{1}{k^2}\nonumber\\
& = & (1-{\rm sgn}(u))\sum_{k\,{\rm even}}\frac{1}{k^2}+{\rm sgn(u)}\sum_{k=1}^\infty\frac{1}{k^2}\nonumber\\
& = & \frac{1}{4}(1+3\,{\rm sgn(u)})\sum_{k=1}^\infty\frac{1}{k^2}=\frac{\pi^2}{24}+{\rm sgn}(u)\frac{\pi^2}{8}\,.
\eeq
Moreover, depending on the sign of $u$, we can split the integral into an integral from $0$ to ${\rm sgn}(u)$ (which gives $\Phi({\rm sgn}(u))$) and an integral from ${\rm sgn}(u)$ to $u$ on which we implement the change of variables $x=1/y$. We find
\beq
\Phi(u) & =  & \Phi({\rm sgn}(u))+\int_{{\rm sgn}(u)}^{1/u}\frac{dy}{y}\Big[\ln|1-y|-\ln |y|\Big]\nonumber\\
& = & 2\Phi({\rm sgn}(u))-\Phi\left(\frac{1}{u}\right)-\frac{1}{2}\ln^2|u|\,,
\eeq
that is
\beq
\Phi(u)+\Phi\left(\frac{1}{u}\right)=\frac{\pi^2}{12}+{\rm sgn}(u)\frac{\pi^2}{4}-\frac{1}{2}\ln^2|u|\,.
\eeq
Using these various formulas, together with $t_\tau t_{-\tau}=1$, we arrive at
\beq\label{eq:idddd}
\int_{\sigma}^{t_\tau}dt\,\frac{\ln (ix)}{t} & = & \frac{\pi^2}{12}-\sigma\frac{\pi^2}{4}-2 \Phi (-t_\tau)+\ln m^2 \ln |t_\tau|-\frac{1}{2}\ln^2|t_\tau| + i \frac{\pi}{2}\sigma\ln |t_\tau|\nonumber\\
& = & -\left[\frac{\pi^2}{12}-\sigma\frac{\pi^2}{4}-2 \Phi (-t_{-\tau})+\ln m^2 \ln |t_{-\tau}|-\frac{1}{2}\ln^2|t_{-\tau}| + i \frac{\pi}{2}\sigma\ln |t_{-\tau}|\right],\nonumber\\
\eeq
where the second equality is the explicit form of the identity
\beq
\int_{\sigma}^{t_{-\tau}}dt\,\frac{\ln (ix)}{t}=-\int_{\sigma}^{t_{\tau}}dt\,\frac{\ln (ix)}{t}\,,
\eeq
which is readily obtained using the change of variables $t\to 1/t$ and $t_\tau=-1/t_{-\tau}$. This formula will be useful below when checking that our final result does not depend on the choice of $\tau$.

%%%
\subsection{Recursive determination of the $\bar V^{(j)}$'s}
Let us now determine the $\bar V^{(j)}$'s recursively. We start from Eq.~(\ref{eq:gm2}) which gives
\beq
\frac{\partial\bar{g}^{(-2)}(m^2,x)}{\partial x}=\frac{i}{2}m^2\,.
\eeq
Using Eq.~(\ref{eq:int_final}), this leads then to
\beq
\bar V^{(-2)}(m^2,c^2) & = & \frac{i}{8m^2}\Big[\tau m^2R(m^2,c^2)I_0+(2m^2-c^2)(c^2-m^2)\Big]\nonumber\\
& = & \frac{i}{8m^2}\Big[R^2(m^2,c^2)-(c^4-3m^2c^2+2m^4)\Big]\nonumber\\
& = & -\frac{i}{8}(2m^2+c^2)\,.\label{V2}
\eeq
From this result and Eq.~(\ref{eq:gm1}), we find 
\beq
\bar g^{(-1)}(c^2)=\frac{i}{8} \Big[6m^4-2m^2c^2-c^4+2m^2(c^2-2 m^2) \ln (i m^2)+2 m^2c^2\ln (i c^2)\Big],
\eeq
and then
\beq
\frac{\partial\bar{g}^{(-1)}(m^2,x)}{\partial x}= \frac{i}{4}\Big[m^2 \ln(i m^2)-x+ m^2 \ln (i x)\Big].
\eeq
Equation (\ref{eq:int_final}) now gives
\beq
\bar V^{(-1)}(m^2,c^2)&=& \frac{i}{16m^2}\Bigg[ \tau R(m^2,c^2) \left(m^2 \ln (im^2)I_0-I_1+m^2J_0\right)\nonumber\\
& & \hspace{4.0cm}-\,4i\left(2m^2-c^2\right)\frac{\bar g^{(-1)}(m^2,c^2)}{m^2}\Bigg]\nonumber\\
&=&\frac{i}{16}\Big[ 3(2m^2+c^2)-4m^2 \ln (im^2)-2c^2 \ln (i c^2)\Big].
\label{V1}
\eeq
From this result and Eq.~(\ref{eq:gm0}), we find 
\beq
\bar g^{(0)}(m^2,c^2) & = & \frac{i}{16}\Big[-14 m^4+2 m^2 c^2+3c^4+m^2(c^2-4 m^2) \ln^2(i m^2)+m^2 c^2\ln^2(i c^2)\nonumber\\
&&\hspace{0.5cm}+\,2 m^2c^2 \ln (i m^2)\ln (ic^2)+6m^2(2 m^2-c^2)\ln (i m^2)+2c^2(m^2-c^2) \ln (i c^2)\Big]\nonumber\\
\eeq
and then
\beq
\frac{\partial \bar{g}^{(0)}}{\partial x}&=&\frac{i}{16}  \Big[4(m^2+x)-4 m^2 \ln (i m^2)+4 \left(m^2-x\right) \ln (i x)\nonumber\\
& & \hspace{0.5cm}+\,m^2 \ln ^2(i x)+m^2 \ln ^2(i m^2)+2 m^2 \ln (i m^2)\ln (i x)\Big].
\eeq
Equation (\ref{eq:int_final}) now gives
\beq
\bar V^{(0)}(m^2,c^2)&=& \frac{i}{64m^2}\Bigg[ \tau R(m^2,c^2) \Big\{\big(4-4\ln(im^2)+\ln^2(im^2)\big)m^2I_0+4I_1\nonumber\\
& & \hspace{3.5cm}+\,2\big(2+\ln(im^2)\big)m^2 J_0-4J_1+2m^2K_0\Big\}\nonumber\\
& & \hspace{5.5cm}-\,16i\left(2m^2-c^2\right)\frac{\bar g^{(0)}(m^2,c^2)}{m^2}\Bigg]
\label{V0}
\eeq
wich simplifies to 
\beq
\bar V^{(0)}(m^2,c^2) &=&\frac{i}{16}\Bigg[\tau R\left(m^2,c^2\right) \left\{-\frac{\pi ^2}{12}+{\rm sgn}(c^2)\frac{\pi^2}{4}+2 \Phi\left(-t_\tau(c^2)\right)\right.\nonumber\\
& & \hspace{3.2cm}\left.+\,\frac{1}{2} \ln ^2|t_\tau(c^2)|+i \pi \Theta(-c^2)\ln | t_\tau(c^2)|\right\}\nonumber\\
&& \hspace{0.7cm}-\,\frac{7}{2} (2 m^2+c^2)+6 m^2 \ln(i m^2)+3c^2 \ln(i c^2)\nonumber\\
&& \hspace{0.7cm}-\,c^2 \ln(i m^2) \ln (i c^2)+\left(\frac{c^2}{2}-2m^2\right) \ln ^2(i m^2)-\frac{1}{2}c^2 \ln ^2(i c^2)\Bigg],
\label{im}
\eeq
where we recall that
\beq
t_\tau(c^2)=\frac{c^2-2m^2+\tau\sqrt{c^2(c^2-4m^2)}}{2m^2}\,.
\eeq
It is clear from (\ref{eq:idddd}), that this result does not depend on $\tau$. In particular, it is convenient to choose $\tau=+1$ since $-t_+<1$ and therefore $\Phi(-t_+)={\rm Li}_2(-t_+)$:
\beq
\bar V^{(0)}(m^2,c^2) &=&\frac{i}{16}\Bigg[R\left(m^2,c^2\right) \left\{-\frac{\pi ^2}{12}+{\rm sgn}(c^2)\frac{\pi^2}{4}+2 {\rm Li}_2\left(-t_+(c^2)\right)\right.\nonumber\\
& & \hspace{3.2cm}\left.+\,\frac{1}{2} \ln^2|t_+(c^2)|+i \pi \Theta(-c^2)\ln |t_+(c^2)|\right\}\nonumber\\
&& \hspace{0.7cm}-\,\frac{7}{2} (2 m^2+c^2)+6 m^2 \ln(i m^2)+3c^2 \ln(i c^2)\nonumber\\
&& \hspace{0.7cm}-\,c^2 \ln(i m^2) \ln (i c^2)+\left(\frac{c^2}{2}-2m^2\right) \ln ^2(i m^2)-\frac{1}{2}c^2 \ln ^2(i c^2)\Bigg].
\label{im2}
\eeq
For the relevant case $c^2=-m^2$, we have $t_+(-m^2)=(-3+\sqrt{5})/2$. Using the well known result
\beq
{\rm Li}_2\left(\frac{3-\sqrt{5}}{2}\right)=\frac{\pi^2}{15}-\ln^2\frac{1+\sqrt{5}}{2}\,,
\eeq
as well as
\beq
\ln\frac{3-\sqrt{5}}{2}=-2\ln\frac{1+\sqrt{5}}{2}\,,
\eeq
we find
\beq
& & \bar V^{(0)}(m^2,-m^2)=-\frac{im^2}{16}\Bigg[\sqrt{5}\left(\frac{\pi ^2}{5}-i \pi\ln \frac{3-\sqrt{5}}{2}\right)+\frac{7}{2}\nonumber\\
&& \hspace{2.5cm}-\,6 \ln(i m^2)+3\ln(-i m^2)-\ln(i m^2) \ln (-i m^2)+\frac{5}{2} \ln ^2(i m^2)-\frac{1}{2} \ln ^2(-i m^2)\Bigg].\nonumber\\\label{V0}
\eeq
 Combining Eqs.~\eqref{V2}, \eqref{V1} and \eqref{V0} into \eqref{eq:laporta}, we arrive eventually at
\beq
& & S_{\alpha \alpha(-\alpha)}(0n)\nonumber\\
& & \hspace{0.5cm}=\,(4\pi\mu^2)^{2\epsilon}\Gamma\left(1+\epsilon\right)^2\left(-\frac{\alpha}{128\pi^4}\right)\Bigg[\frac{1}{4\epsilon^2}+\frac{1}{2\epsilon}\left(\frac{3}{2}-2 \ln \alpha+\ln (-\alpha)\right)\nonumber\\
& & \hspace{2.5cm}+\,\frac{\sqrt{5}}{2}\left(\frac{\pi ^2}{5}-i \pi\ln \frac{3-\sqrt{5}}{2}\right)\nonumber\\
&& \hspace{2.5cm}+\,\frac{7}{4}-3 \ln\alpha+\frac{3}{2}\ln(-\alpha)-\frac{1}{2}\ln\alpha \ln (-\alpha)+\frac{5}{4} \ln ^2 \alpha-\frac{1}{4} \ln ^2(-\alpha)\Bigg].
\eeq

\end{document}